\documentclass{vldb}
\usepackage{times}
\usepackage{graphicx}
\usepackage{fancyvrb}
\usepackage{color}
\usepackage{scalops-listings}
\usepackage{url}
\usepackage{xspace}
\usepackage{ntheorem}
\usepackage{todonotes}
\usepackage{multirow}
\usepackage{pgfplots}
\usepackage{subfigure}

\newcommand{\eat}[1]{}

\title{Scaling Datalog for Machine Learning on Big Data}

\numberofauthors{3} 

\author{
\alignauthor {\text{Yingyi Bu, Vinayak Borkar, Michael J. Carey}} \\
\affaddr{University of California, Irvine}
\alignauthor {\text{Joshua Rosen, Neoklis Polyzotis}} \\
\affaddr{University of California, Santa Cruz}
\and  
\alignauthor {\text{Tyson Condie, Markus Weimer, Raghu Ramakrishnan}} \\
\affaddr{Yahoo! Research}
}

\newtheorem{dfn}{Definition}
\newtheorem{thm}{Theorem}

\newcommand{\ol}[1]{\texttt{\small #1}\xspace}

\newcommand{\inner}[2]{\left\langle #1,#2 \right\rangle}

\newcommand{\argmin}[1]{\underset{#1}{\operatorname{argmin}}}

\makeatletter
\let\@copyrightspace\relax
\makeatother

\begin{document}
\maketitle

\sloppy
\begin{abstract}
In this paper, we present the case for a declarative foundation for
data-intensive machine learning systems.  Instead of creating a new system for
each specific flavor of machine learning task, or hardcoding new optimizations,
we argue for the use of {\it recursive queries} to program a variety of machine
learning systems.  By taking this approach, database query optimization
techniques can be utilized to identify effective execution plans, and the
resulting runtime plans can be executed on a single {\it unified} data-parallel
query processing engine.  As a proof of concept, we consider two programming
models---Pregel and Iterative Map-Reduce-Update---from the machine learning
domain, and show how they can be captured in Datalog, tuned for a specific
task, and then compiled into an optimized physical plan.  Experiments performed
on a large computing cluster with real data demonstrate that this declarative
approach can provide very good performance while offering both increased
generality and programming ease.

\end{abstract}

\makeatletter
\def\topfigrule{\kern3\p@ \hrule \kern -3.4\p@} 
\def\botfigrule{\kern-3\p@ \hrule \kern 2.6\p@} 
\def\dblfigrule{\kern3\p@ \hrule \kern -3.4\p@} 
\makeatother \addtolength{\textfloatsep}{-.5\textfloatsep}
\addtolength{\dbltextfloatsep}{-.5\dbltextfloatsep}
\addtolength{\floatsep}{-.5\floatsep}
\addtolength{\dblfloatsep}{-.5\dblfloatsep}

\section{Introduction}\label{sec:introduction}

Supported by the proliferation of ``Big Data'' platforms such as Hadoop,
organizations are collecting and analyzing ever larger datasets.  Increasingly,
machine learning (ML) is at the core of data analysis for actionable business
insights and optimizations.  Today, machine learning is deployed widely:
recommender systems drive the sales of most online shops; classifiers help keep
spam out of our email accounts; computational advertising systems drive
revenues; content recommenders provide targeted user experiences;
machine-learned models suggest new friends, new jobs, and a variety of
activities relevant to our profile in social networks.  Machine learning is
also enabling scientists to interpret and draw new insights from massive
datasets in many domains, including such fields as astronomy, high-energy
physics, and computational biology.

The availability of powerful distributed data platforms and the widespread
success of machine learning has led to a virtuous cycle wherein organizations
are now investing in gathering a wider range of (even bigger!) datasets and
addressing an even broader range of tasks.  Unfortunately, the basic MapReduce
framework commonly provided by first-generation ``Big Data analytics''
platforms like Hadoop lacks an essential feature for machine learning:
MapReduce does not support iteration (or equivalently, recursion) or certain
key features required to efficiently iterate ``around'' a MapReduce program.
Programmers building ML models on such systems are forced to implement looping
in ad-hoc ways outside the core MapReduce framework; this makes their
programming task much harder, and it often also yields inefficient programs in
the end.  This lack of support has motivated the recent development of various
specialized approaches or libraries to support iterative programming on large
clusters.  Examples include Pregel, Spark, and Mahout, each of which aims to support
a particular family of tasks, e.g., graph analysis or certain types of ML models,
efficiently.  Meanwhile, recent MapReduce extensions such as HaLoop, Twister,
and PrItr aim at directly addressing the iteration outage in MapReduce; they do
so at the physical level, however.

The current generation of specialized platforms seek to improve a user's
programming experience by making it much easier (relative to MapReduce) to
express certain classes of parallel algorithms to solve ML and graph analytics
problems over Big Data. Pregel is a prototypical example of such a
platform; it allows problem-solvers to ``think like a vertex'' by writing a few
user-defined functions (UDFs) that operate on vertices, which the framework can
then apply to an arbitrarily large graph in a parallel fashion.  Unfortunately
for both their implementors and users, each such platform is a distinct new
system that has been built from the ground up.  Ideally, a specialized platform
should allow for better optimization strategies for the class of problems
considered ``in scope.'' In reality, however, each new system is built from
scratch and must include efficient components to perform common tasks such as
scheduling and message-passing between the machines in a cluster.  Also, for
Big Data problems involving multiple ML algorithms, it is often necessary
to somehow glue together multiple platforms and to pass (and translate) data
via files from one platform to another.  It would clearly be attractive if
there were a common, general-purpose platform for data-intensive computing
available that could simultaneously support the required programming models and
allow various domain-specific systems the ability to reuse the common pieces.
Also desirable would be a much cleaner separation between the logical
specification of a problem's solution and the physical runtime strategy to be
employed; this would allow alternative runtime strategies to be considered for
execution, thus leading to more efficient executions of different sorts of 
jobs.

In this paper, we show that it is indeed possible to provide a declarative
framework capable of efficiently supporting a broad range of machine learning
and other tasks that require iteration, and then to develop specialized
programming models that target specific classes of ML tasks on top of this
framework.  Hence, much of the effort that is currently involved in building
such specialized systems is factored out into a single underlying optimizer and
runtime system.  We propose the use of Datalog, which allows recursive queries
to be naturally specified, as the common declarative language ``under the
hood'', into which we map high-level programming models.~\footnote{We
leave open the possibility of exposing Datalog as an ``above the hood''
user-programmable language; doing so would place a premium upon being able
to optimize arbitrary Datalog programs in a ``Big Data'' cluster environment,
which our current results do not yet
address.} Moreover, Datalog can readily express the {\em dataflow} aspects of a
ML program, which are the main drivers of cost in a Big Data setting.  Our
approach opens the door to applying the rich body of work on optimizing Datalog
programs and identifying effective execution plans, and allows us to execute
the resulting plans on a single {\it unified} data-parallel query processing
engine over a large cluster.

\begin{figure}
  \begin{center}
   \includegraphics[scale=0.35]{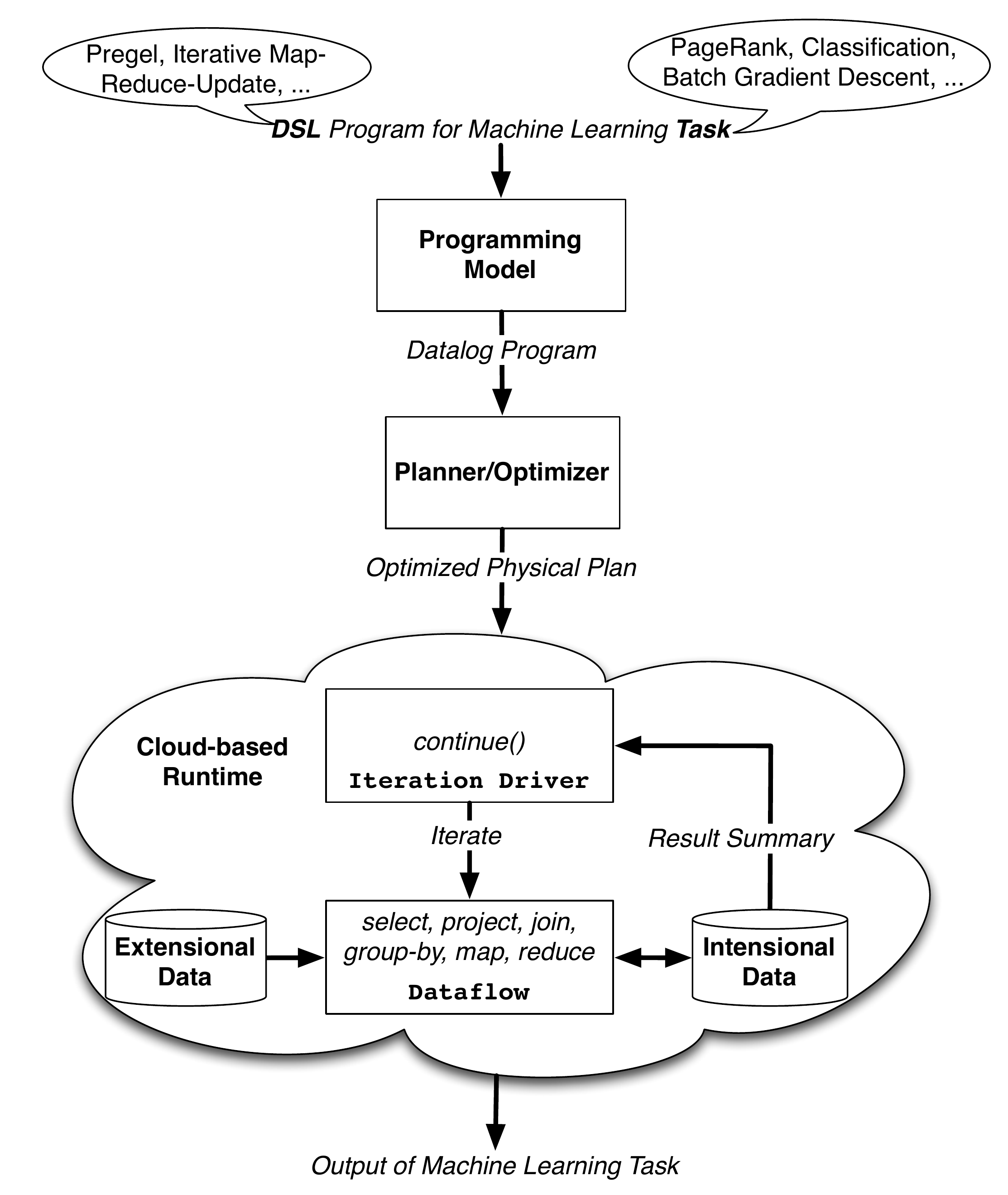}
   \caption{\label{fig:architecture}System stack for large-scale Machine Learning.}
  \end{center}
\end{figure}

Figure~\ref{fig:architecture} sketches the approach advocated here.  A
domain-specific programming model---such as Pregel or Iterative
Map-Reduce-Update---is used to program a specific ML task, e.g.,
PageRank~\cite{pagerank} or Batch Gradient Descent.  The task program is then
translated to a Datalog program that captures the intended programming model as
declarative rule specifications and the task-specific code (e.g., the PageRank
algorithm) as UDFs.  Subsequently, a planner/optimizer compiles the declarative
Datalog program into an efficient execution plan.  Lastly, a cloud-based
runtime engine---consisting of a dataflow of data-parallel operators,
extensional and intensional datasets, and an iteration driver---executes the
optimized plan to a fixed point, producing the output of the ML task.  Central
to our thesis is that by capturing the ML programming model in a high-level
declarative language, we can automatically generate physical plans that are
optimized---based on hardware configurations and data statistics---for a target
class of ML tasks.

As a concrete example of the benefits of our approach,consider Pregel
again, a specialized programming model and runtime tuned to
graph-oriented computations.  Suppose a data scientist wants to train
a model through a Batch Gradient Descent (BGD) task using Pregel,
which requires them to ``think like a vertex.'' A possible approach
would be to encode each data point as a vertex that computes a
(gradient, loss) value and sends it to a global aggregator, which sums
all the values to a global statistic and updates the model.  This
process repeats over a series of ``supersteps'' until the model
converges to some value.  There are two problems with this approach.
Firstly, it is unnatural to treat the training data---a set of
unrelated feature vectors---as a graph.  Secondly, encoding a BGD task
into a general purpose graph-oriented programming model is suboptimal,
in terms of programming ease and runtime performance.  For more
appropriate graph-analysis tasks, like PageRank, the Pregel
programming model and runtime contains hardcoded features---such as
non-monotonic halting conditions---that are often not required.  The
net result is that Pregel is suitable only for a specific class of ML
tasks, and not appropriate or suboptimal for others.  In contrast, we
propose capturing an ML task as a declarative Datalog program, and
then letting a query optimizer translate it to an appropriate physical
plan.

To summarize, in our proposed approach to scalable machine learning, programmers
need not learn to operate and use a plethora of distinct platforms. Relational
database systems separate the conceptual, logical and physical
schemas in order to achieve {\em logical} and {\em physical data independence}.
Similarly, we open the door
to principled optimizations by achieving a separation between:
\begin{list}{\labelitemi}{\leftmargin=1em}\itemsep 0pt \parskip 0pt
\item
The user's program (in a sublanguage of their choice, making
use of a library of available templates and user-definable functions)
and the underlying {\em logical query}, expressed in Datalog. This
shields the user from any changes in the logical framework, e.g., 
how the Datalog program is optimized.
\item
The logical Datalog query and an optimized {\em physical runtime plan},
reflecting details related to caching, storage, indexing, the logic for
incremental evaluation and re-execution in the face of failures, etc.
This ensures that any enhancements to the plan execution 
engine will automatically translate to more efficient execution, without
requiring users to re-program their tasks to take advantage of the enhancements.
\end{list}

In essence, the separation identifies ``modules'' (such as the plan execution
engine or the optimization of the logical Datalog program) where localized
enhancements lead to higher overall efficiency.  To illustrate our approach, we
will show here how the Pregel and Iterative Map-Reduce-Update programming
models can each be translated into Datalog programs.  Second, we will
demonstrate that an appropriately chosen data-intensive computing substrate,
namely Hyracks~\cite{Borkar:2011ly}, is able to handle the computational
requirements of such programs through the application of dataflow processing
techniques like those used in parallel databases~\cite{DeWitt:1992}.  This
demonstration involves the presentation of experimental results obtained by
running the sorts of Hyracks jobs that will result from our translation stack
against real data on a large computational cluster at Yahoo!\@.  Our findings
indicate that such a declarative approach can indeed provide very good
performance while offering increased generality and ease of programming.

The remainder of this paper is organized as follows:
Section~\ref{sec:programming-models} provides a brief data-centric perspective
on machine learning and then reviews two popular programming models in use
today for scalable machine learning.  Section~\ref{sec:declarative} shows how
programs written against these two programming models can be captured in
Datalog and then translated into an extended relational algebra, while
Section~\ref{sec:physical} describes the resulting physical plans.
Section~\ref{sec:experiments} presents preliminary experimental results
obtained by running the physical Hyracks plans for two tasks---Batch Gradient
Descent and PageRank---on a large research cluster at Yahoo!\@ against real
datasets.  Section~\ref{sec:related} relates this work to other activities in
this area, past and present, and Section~\ref{sec:conclusion} presents our
conclusions.

\section{Programming Models for ML}\label{sec:programming-models}
The goal of machine learning (ML) is to turn observational data into a
\emph{model} that can be used to predict for or explain yet unseen
data.  While the range of machine learning techniques is broad, most
can be understood in terms of three complementary perspectives:
\begin{list}{\labelitemi}{\leftmargin=1em}\itemsep 0pt \parskip 0pt
\item {\bf ML as Search:} The process of training a model can be
  viewed as a \emph{search problem}.  A domain expert writes a program
  with an objective function that contains, possibly millions of,
  unknown parameters, which together form the {\em model}.  A runtime
  program {\em searches} for a good set of parameters based on the
  objective function.  A {\em search strategy} enumerates the
  parameter space to find a model that can correctly capture the known
  data and accurately predict unknown instances.
\item {\bf ML as Iterative Refinement:} Training a model can be viewed
  as iteratively closing the gap between the model and underlying
  reality being modeled, necessitating iterative/recursive
  programming.
\item {\bf ML as Graph Computation:} Often, the interdependence
  between the model parameters is expressible as a graph, where nodes
  are the parameters (e.g., statistical/random variables) and edges
  encode interdependence.  This view gives rise to the notion of
  \emph{graphical models}, and algorithms often reflect the graph
  structure closely (e.g., propagating refinements of parameter
  estimates iteratively along the edges, aggregating the inputs at
  each vertex).
\end{list}
Interestingly, each of these perspectives lends itself readily to
expression as a Datalog program, and as we argue in this paper,
thereby to efficient execution by applying a rich array of
optimization techniques from the database literature.  The fact that
Datalog is well-suited for iterative computations and for
graph-centric programming is well-known~\cite{deductive-database}, and
it has also been demonstrated that Datalog is well-suited to search
problems~\cite{evitaraced}.  The natural fit between Datalog and ML
programming has also been recognized by others~\cite{Atul-thesis,
  eisner-filardo-2011}, but not at ``Big Data'' scale.  It is our goal
to make the optimizations and theory behind Datalog available to
large-scale machine learning while facilitating the use of established
programming models.  To that end, we advocate \emph{compilation} from
higher order programming models to Datalog and subsequently physical
execution plans.

To study the feasibility of this approach, we show how two major
programming models supporting distributed machine learning
today---Pregel and Iterative Map-Reduce-Update---can be expressed in
Datalog and subsequently efficiently executed.  In the remainder of
this Section, we describe these two programming models in more detail
with the goal of isolating the user code in terms of user-defined
functions (UDFs).  This will set the stage for the next section, in
which we present concise Datalog representations for these two ML
programming models, reusing the UDFs introduced here.  As we will then
see, the structure of many ML problems is inherently recursive.

\subsection{Pregel}\label{sec:pregel}
Pregel~\cite{pregel} is a system developed at Google for supporting
graph analytics.  It exposes a message-passing interface in the form
of two per-vertex UDFs:
\begin{description}
\item[\ol{update}] the per-vertex update function.  It accepts the
  current vertex state and inbound messages and produces outbound
  messages as well as an updated state.
\item[\ol{combine} (optional)] aggregates messages destined for a
  vertex.
\end{description}
We omit other aspects of Pregel---graph mutation and global
aggregators---because they are not necessary in many graph
algorithms~\cite{pregel}, and the machinery for global aggregators is
captured later when we address Iterative Map-Reduce-Update.

The Pregel runtime executes a sequence of iterations called {\em
  supersteps} through a bulk-synchronous processing (BSP) model.  In a
single superstep, the Pregel runtime executes the \ol{update} UDF on
all \emph{active} vertices exactly once.  A vertex is active in the
current superstep if there are messages destined for it or if the
\ol{update} UDF indicates---in the previous superstep---its desire to
execute.  The output of a superstep can be materialized for fault
tolerance before executing the subsequent superstep.  The runtime
halts when no vertices are active.

\textbf{Example: PageRank}~\cite{pagerank} is a canonical example of a
graph algorithm that is concisely captured by Pregel.  Websites are
represented by vertices and hyperlinks form the edges of the graph.
In a single superstep, the \ol{update} UDF receives the PageRank of
the current vertex and its neighbors.  It emits the updated PageRank
for this vertex and if its PageRank changed sufficiently, its new
value is sent to its neighbors.  The system converges when no more
such updates occur or a maximum number of supersteps is reached.

\subsection{Iterative Map-Reduce-Update}\label{sec:itermr}
A large class of machine learning algorithms are expressible in the
statistical query model~\cite{kearns93t}. Statistical queries
(e.g. max, min, sum, \dots) themselves decompose into a data-local
\ol{map} function and a subsequent aggregation using a \ol{reduce}
function~\cite{Chu:2006fk}, where \ol{map} and \ol{reduce} refer to
the functions by the same name from the functional programming
literature.  We support the resulting Iterative Map-Reduce-Update
programming model through the following three UDFs:
\begin{description}
\item[\ol{map}] receives read-only global state as side information
  and is applied to all training data points in parallel.
\item[\ol{reduce}] aggregates the \ol{map}-output. This function is
  commutative and associative.
\item[\ol{update}] receives the combined aggregated results and
  produces a new global state for the next iteration or indicates that
  no additional iteration is necessary.
\end{description}
An Iterative Map-Reduce-Update runtime executes a series of
iterations, each of which first calls \ol{map} with the required
arguments, then performs a global \ol{reduce} aggregation, and lastly
makes a call to \ol{update}.  We assume the runtime terminates when
\ol{update} returns the same model that it was
given.~\footnote{Alternatively, a vote to halt protocol could be
  simulated through a boolean value in the model that signals the
  termination.} It is interesting to point out here that Google's
MapReduce~\cite{Dean:2004uq} programming model is not an ideal fit: it
contains a group-by key component that is not needed for many
statistical queries from the machine learning domain.  Additionally,
we require an iterative looping construct and an update step that fall
outside the scope of Google's MapReduce framework.

\textbf{Example: Convex Optimization} A large class of machine
learning---including Support Vector Machines, Linear and Logistic
Regression and structured prediction tasks such as machine
translation---can be cast as convex optimization problems, which in
turn can be solved efficiently using an Iterative Map-Reduce-Update
approach~\cite{Alekh-Agarwal:2011fk, Weimer:2010fk}.  The objective is
to minimize the sum over all data points of the divergences (the loss)
between the model's prediction and the known data.  Usually, the loss
function is convex and differentiable in the model, and therefore the
\emph{gradient} of the loss function can be used in iterative
optimization algorithms such as Batch Gradient Descent.~\footnote{A
  more detailed discussion can be found in the Appendix~\ref{sec:bgd}.} Each model update step is a
single Map-Reduce-Update iteration.  The \ol{map} UDF computes (loss,
gradient) tuples for all data points, using the current model as side
input.  The \ol{reduce} UDF sums those up and \ol{update} updates the
model.  The updated model becomes the input of the next iteration of
the optimization algorithm.

\subsection{Discussion}

From a data flow perspective, a key differentiator between different
types of ML models is the relationship of the model to the
observational data; both in size and structure.  Certain models (e.g.,
regression, classification and clustering) are \emph{global} to all
observation data points and are relatively small in size (think MB vs.
GB).  In others (e.g., topic models and matrix factorization), the
model consists of interdependent parameters that are \emph{local} to
each observation and are therefore on the same order-of-magnitude in
terms of size as the observational data.  Any system that seeks to
support both classes of ML models efficiently must recognize the
nature of the task---global or local---and be able to optimize
accordingly.

The two programming frameworks that we consider span a wide area of
machine learning and graph analytics.  Pregel is a well-known graph
analytics platform that can be used to develop local models.
Iterative Map-Reduce-Update is gaining traction as an ideal framework
for producing global models.  It is important to point out that both
frameworks can express each other---Pregel can be implemented on top
of an Iterative Map-Reduce-Update system and vice versa---but that
each system was designed and optimized for a specific ``native''
application type.  Abusing one to solve the other would incur
significant performance overheads.  Instead, in
Section~\ref{sec:declarative}, we unify these frameworks and the
intended semantics as declarative specifications written in the
Datalog language.  We then show a direct translation from the Datalog
specifications to a data-parallel recursive runtime that is able to
retain the performance gains offered by runtimes specifically tuned to
a given framework.

\section{Declarative Representation}\label{sec:declarative}

This section presents a translation of the two programming models into
declarative Datalog programs and a formal analysis of the semantics
and correctness of this translation.  In doing so, we expose
information about the semantics and structure of the underlying data
operations, which in turn allows us to reason about possible
optimizations (e.g., reordering the operators or mapping logical
operators to different possible implementations) and thus generate
efficient execution plans over a large range of configurations.
Datalog is a natural choice for this intermediate logical
representation, as it can encode succinctly the inherent recursion of
the algorithms.




Before diving into the details of the translation of the two
programming models, we present a short overview of the main concepts
in Datalog.  A Datalog program consists of a set of {\em rules} and an
optional {\em query}.  A Datalog rule has the form
\ol{$p(\mathbf{Y})$~:-~$q_1$($\mathbf{X_1}$),~$\ldots$,~$q_n$($\mathbf{X_n}$)},
where $p$ is the head predicate of the rule, $q_1,\dots,q_n$ are
called the body predicates, and
$\mathbf{Y},\mathbf{X_1},\dots,\mathbf{X_n}$ correspond to lists of
variables and constants.  Informally, a Datalog rule reads ``if there
is an assignment of values
$\mathbf{v},\mathbf{v_1},\dots,\mathbf{v_n}$ corresponding to
$\mathbf{Y},\mathbf{X_1},\dots,\mathbf{X_n}$ such that
$q_1(\mathbf{v_1}) \land\dots\land q_n(\mathbf{v_n})$ is true {\bf
  then} $p(\mathbf{v})$ is true.'' In the rules that we consider, a
predicate can be one of three types:
\begin{list} {\labelitemi}{\leftmargin=1em}\itemsep 0pt \parskip 0pt
\item An extensional predicate, which maps to the tuples of an
  existing relation.  An extensional predicate $q_i(\mathbf{v})$ is
  true if and only if the tuple $\mathbf{v}$ is present in the
  corresponding relation.

\item An intensional predicate, which corresponds to the head $p$ of a
  rule.  Intensional predicates essentially correspond to views.

\item A function predicate, which corresponds to the application of a
  function.  As an example, consider a function $f$ that receives as
  input three datums and outputs a tuple of two datums, and assume
  that the corresponding function predicate is $q_f$.  We say that
  $q_f(v_1,v_2,v_3,v_4,v_5)$ is true if and only if the result of
  $f(v_1,v_2,v_3)$ is $(v_4,v_5)$.  By convention, we will always
  designate the first attributes of the predicate as the inputs to the
  function and the remaining attributes as the output.
\end{list}

We allow group-by aggregation in the head in the form
\ol{$p$(Y,~aggr<Z>)}.  As an example, the rule
\ol{$p$(Y,~SUM<Z>)~:-~$q_1$(Y,~Z)} will compute the sum of $Z$ values
in $q_1$ grouped-by $Y$.  We also allow variables to take set values
and provide a mechanism for member iteration.  As an example, the rule
\ol{$p$(X,~Y)~:-~$q_1$(X,~\{Y\})} implies that the second attribute of
$q_1$ takes a set value, and binds $Y$ to every member of the set in
turn (generating a tuple in $p$ per member, essentially unnesting the
set).

Recursion in Datalog is expressed by rules that refer to each other in
a cyclic fashion.  The order that the rules are defined in a program
is semantically immaterial.  Program evaluation proceeds bottom-up,
starting from the extensional predicates and inferring new facts
through intensional and function predicates.  The evaluation of a
Datalog program reaches a {\em fixpoint} when no further deductions
can be made based on the currently inferred
facts~\cite{deductive-database}.

\eat{
In the remainder of this section, we describe the two programming frameworks
from Section~\ref{sec:programming-models} in Datalog.
Section~\ref{sec:pregel-datalog} discusses the Pregel framework as it pertains
to deriving local models.  In Section~\ref{sec:iterMR-datalog}, we present our
Datalog rules for Iterative Map-Reduce-Update in the context of deriving global
models.  We conclude with a discussion (Section~\ref{sec:discussion-datalog})
on the semantics of our Datalog programs and argue their correctness
relative to the target frameworks.
}

\subsection{Pregel for Local Models}\label{sec:pregel-datalog}

We begin with the Datalog program in Listing~\ref{lst:datalog-local},
which specifies the Pregel programming model as it pertains to
deriving local models.  A special temporal argument (the variable
\ol{J}) is used to track the current superstep number, which will be
passed to the \ol{update} UDF invocation in Rule~\emph{L6} discussed
below.  Rule~\emph{L1} invokes an initialization UDF \ol{init\_vertex}
(which accepts the (\ol{Id, Datum}) variables as argument and returns
the (\ol{State}) variable) on every tuple in the input: referenced by
the \ol{data} predicate.  Rule~\emph{L2} then initializes a \ol{send}
predicate with an activation message to be delivered to all vertices
in iteration zero.  Rule~\emph{L3} implements the combination of
messages that are destined for the same vertex.  It performs a
group-by aggregation over predicate \emph{send}, using the
\emph{combine} aggregate function (which is itself a proxy for
\ol{combine}, explained in Section~\ref{sec:pregel}).  In Pregel, a
vertex may forgo updating the state of a given vertex or global
aggregator for some period of supersteps.  Rules~\emph{L4} and
\emph{L5} maintain a view of the most recent vertex state via the
\ol{local} predicate.

\begin{figure}[!ht]
\centering
\vspace{-3ex}
\begin{lstlisting}[language=Prolog, numberblanklines=false,
  label={lst:datalog-local}, caption={Datalog program for the Pregel
    programming model.  The temporal argument is defined by the $J$
    variable.}, frame=single]
% Initialize vertex state
L1: vertex(0, Id, State) :-
  data(Id, Datum),
  init_vertex(Id, Datum, State).

% Initial vertex message to start iteration 0.
L2: send(0, Id, ACTIVATION_MSG) :-
   vertex(0, Id, _).

% Compute and aggregate all messages.
L3: collect(J, Id, combine<Msg>) :-
  send(J, Id, Msg).

% Most recent vertex timestamp
L4: maxVertexJ(Id, max<J>) :-
  vertex(J, Id, State).

% Most recent vertex local state
L5: local(Id, State) :-
  maxVertexJ(Id, J), vertex(J, Id, State).

% new state and outbound messages.
L6: superstep(J, Id, OutState, OutMsgs) :-
  collect(J, Id, InMsgs),
  local(Id, InState),
  update(J,Id,InState,InMsgs,OutState,OutMsgs).

% Update vertex state for next superstep.
L7: vertex(J+1, Id, State) :-
  superstep(J, Id, State, _),
  State != null.

% Flatten messages for the next superstep.
L8: send(J+1, Id, M) :-
  superstep(J, _, _, {(Id,M)}).
\end{lstlisting}
\end{figure}

Rule~\emph{L6} implements the core logic in a superstep by matching
the collected messages with the target local state and then evaluates
the function predicate \emph{update} (which corresponds to UDF
\ol{update}).  The (\ol{J, Id, InState, InMsgs}) variables represent
the arguments and the (\ol{OutState, OutMsgs}) variables hold the
return values: a new state and set of outbound messages.

Finally, rules~\emph{L7} and \emph{L8} stage the next superstep:
\emph{L7} updates the state of each vertex, and \emph{L8} forwards
outbound messages to the corresponding vertices.  Note that the body
of \emph{L7} is conditioned on a non-null state value.  This allows
vertices to forgo state updates in any given superstep.  Finally, the
vote to halt protocol is implemented in the \ol{update} UDF, which
produces a special ``self'' message that activates the vertex in the
next superstep.

\subsection{Iterative Map-Reduce-Update}\label{sec:iterMR-datalog}

\eat{
We now develop the Datalog program for an Iterative Map-Reduce-Update programming model.
\subsubsection{Building Blocks}

We begin with the predicates and functions that form the building
blocks of the Datalog program.
\begin{description}
\item [\emph{training\_data($Id$,$Datum$)}] is again an extensional
  predicate corresponding to the training dataset.
\item [\emph{model($J$,$M$)}] is an intensional predicate recording
  the global model $M$ at iteration $J$. Initialization is performed
  through the function predicate \emph{init\_model($M$)} that returns
  an initial global model.
\item [\emph{map}({\bf M}, {\bf R}, $S$)] is a UDF predicate
  that receives as input the current model $M$ and a data element $R$,
  and generates a statistic $S$ as output.
\item [\emph{reduce}] is an aggregation UDF function that aggregates several per-record statistics
  into one statistic.
\item [\emph{update({\bf J}, {\bf M}, {\bf AggrS}, $NewM$)}] is a UDF
  predicate that receives as input the current iteration $J$, the current model $M$ and the aggregated
  statistic $AggrS$, and generates a new model $NewM$.
\end{description}

\subsubsection{Datalog Program}
}
\begin{figure}
\vspace{-3ex}
\centering
\begin{lstlisting}[language=Prolog, numberblanklines=false,
                   caption={Datalog runtime for the Iterative Map-Reduce-Update programming model.
		   The temporal argument is defined by the $J$ variable.},
		   label={lst:datalog-global},
                   frame=single]
% Initialize the global model
G1: model(0, M) :- init_model(M).

% Compute and aggregate all outbound messages
G2: collect(J, reduce<S>) :- model(J, M),
  training_data(Id, R), map(R, M, S).

% Compute the new model
G3: model(J+1, NewM) :-
   collect(J, AggrS), model(J, M),
   update(J, M, AggrS, NewM), M != NewM.
\end{lstlisting}
\end{figure}

The Datalog program in Listing~\ref{lst:datalog-global} specifies the
Iterative Map-Reduce-Update programming model.  Like before, a special
temporal variable (\ol{J}) is used to track the iteration
number. Rule~\emph{G1} performs initialization of the global model at
iteration~$0$ through function predicate \ol{init\_model}, which takes
no arguments and returns the initial model in the (\ol{M}) variable.
Rules~\emph{G2}~and~\emph{G3} implement the logic of a single
iteration.  Let us consider first rule~\emph{G2}.  The evaluation of
\ol{model(J, M)} and \ol{training\_data(Id, R)} binds (\ol{M}) and
(\ol{R}) to the current global model and a data record respectively.
Subsequently, the evaluation of \ol{map(M, R, S)} invokes the UDF that
generates a data statistic (\ol{S}) based on the input bindings.
Finally, the statistics from all records are aggregated in the head
predicate using the \ol{reduce} UDF (defined in
Section~\ref{sec:itermr}).

Rule~\emph{G3} updates the global data model using the aggregated
statistics.  The first two body predicates simply bind (\ol{M}) to the
current global model and (\ol{AggrS}) to the aggregated statistics
respectively.  The subsequent function predicate \ol{update(J, M,
  AggrS, NewM)} calls the \ol{update} UDF; accepting (\ol{J, M,
  AggrS}) as input and producing an updated global model in the
(\ol{NewM}) variable.  The head predicate records the updated global
model at time-step \ol{J+1}.

Program termination is handled in rule \emph{G3}.  Specifically,
\ol{update} is assumed to return the same model when convergence is
achieved.  In that case, the predicate \ol{M $!=$ NewM} in the body of
\emph{G3} becomes false and we can prove that the program terminates.
Typically, \ol{update} achieves this convergence property by placing a
bound on the number of iterations and/or a threshold on the difference
between the current and the new model.

\subsection{Semantics and Correctness}\label{sec:discussion-datalog}

Up to this point, we have argued informally that the two Datalog
programs faithfully encode the two programming models.  However, this
claim is far from obvious.  The two programs contain recursion that
involves negation and aggregation, and hence we need to show that each
program has a well-defined output.  Subsequently, we have to prove
that this output corresponds to the output of the target programming
models.  In this section, we present a formal analysis of these two
properties.

The foundation for our correctness analysis is based on the following theorem,
which determines that the two Datalog programs fall into the specific class of
XY-stratified programs.

\begin{thm}
  The Datalog programs in Listing~\ref{lst:datalog-local} and
  Listing~\ref{lst:datalog-global} are XY-stratified~\cite{Zaniolo93}.
\end{thm}

\noindent
The proof can be found in Appendix~\ref{sec:semantics} and is based on the
machinery developed in~\cite{Zaniolo93}.  XY-stratified Datalog is a
more general class than stratified Datalog.  In a nutshell, it
includes programs whose evaluation can be stratified based on data
dependencies even though the rules are not stratified.

\eat{
XY-stratification implies
the following key properties for each of our Datalog programs:
\begin{list}{\labelitemi}{\leftmargin=1em}\itemsep 0pt \parskip 0pt
\item The obvious bottom-up method to run the program---starting with the
  extensional facts and recursively inferring additional facts through
  intensional and function predicates until a fixpoint---yields a
  minimal model of the program. This property guarantees that the
  output of the program is well-defined.
\item Fixpoint evaluation can proceed in terms of successive
  iterations, where the input of each iteration is the output of the
  previous iteration. In other words, fixpoint evaluation matches the
  structure of the programming models.
\item Within each stage, each rule of the recursive program needs to
  be fired exactly once. Moreover, rules are fired in a fixed
  sequence. This property enables a highly efficient method to
  evaluate the program. Most importantly, it allows us to reason about
  the order in which UDFs are invoked and hence about the correctness
  of the program.
\end{list}
}

We describe the semantics of the two Datalog programs by translating them
(using standard techniques from the deductive database
literature~\cite{deductive-database}) into an extended relational algebra.  The
resulting description illustrates clearly that the Datalog program encodes
faithfully the corresponding machine-learning task.  Moreover, we can view the
description as a logical plan which can become the input to an optimizing query
processor, which we discuss in the next section.


\begin{figure}
\begin{center}
\includegraphics[scale=0.4]{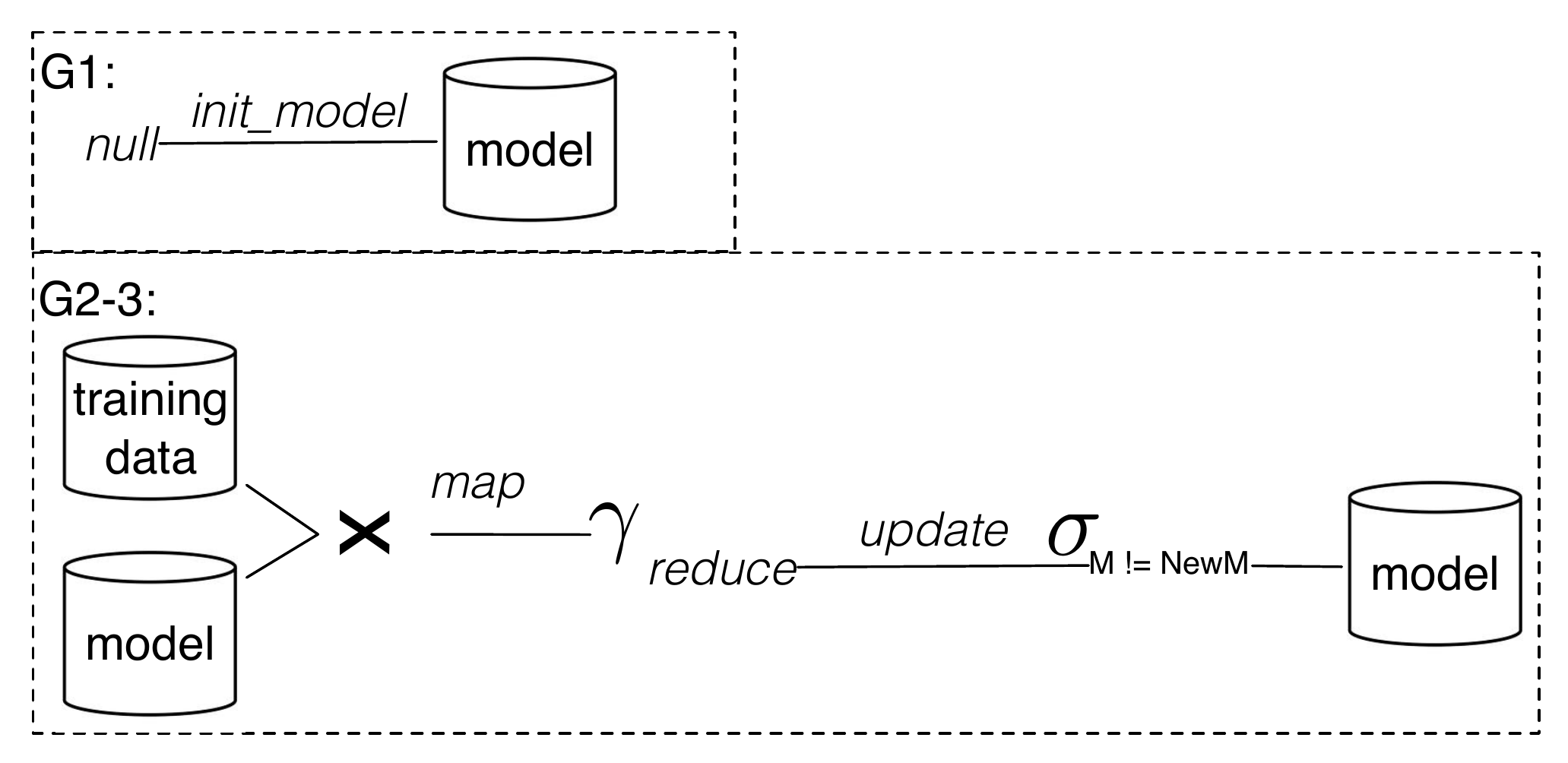}
\caption{\label{fig:bgd-query-plan}Logical query plan for Iterative
  Map-Reduce-Update.}
\end{center}
\end{figure}

To facilitate exposition, we examine first the Datalog program for Iterative
Map-Reduce-Update (Listing~\ref{lst:datalog-global}).  Following
XY-stratification, we can prove that the output of the program is computed from
an initialization step that fires \emph{G1}, followed by several iterations
where each iteration fires \emph{G2} and then \emph{G3}.  By translating the
body of each rule to the corresponding relational algebra expression, and
taking into account the data dependencies between rules, it is straightforward
to arrive at the logical plan shown in Figure~\ref{fig:bgd-query-plan}.  The
plan is divided into two separate dataflows, each labeled by the rules they
implement in Listing~\ref{lst:datalog-global}.  The dataflow labeled $G1$
initializes the global model using the \ol{init\_model} UDF\@, which takes no
input, and produces the initial model.  The $G2\mbox{--}3$ dataflow executes
one iteration.  The \ol{cross-product} operator combines the model with each
tuple in the training dataset, and then calls the \ol{map} UDF on each result.
(This part corresponds to the body of rule \emph{G2}.) The mapped output is
passed to a \ol{group-all} operator, which uses the \ol{reduce} aggregate
(e.g., sum) to produce a scalar value.  (This part corresponds to the head of
rule \emph{G2}.) The aggregate value, together with the model, is then passed
to the \ol{update} UDF\@, the result of which is checked for a new model.
(This part corresponds to the body of \emph{G3}.) A new model triggers a
subsequent iteration, otherwise the \ol{update} output is dropped and the
computation terminates.  (This part corresponds to the head of \emph{G3}.)

\begin{figure}
  \centering
    \includegraphics[scale=0.38]{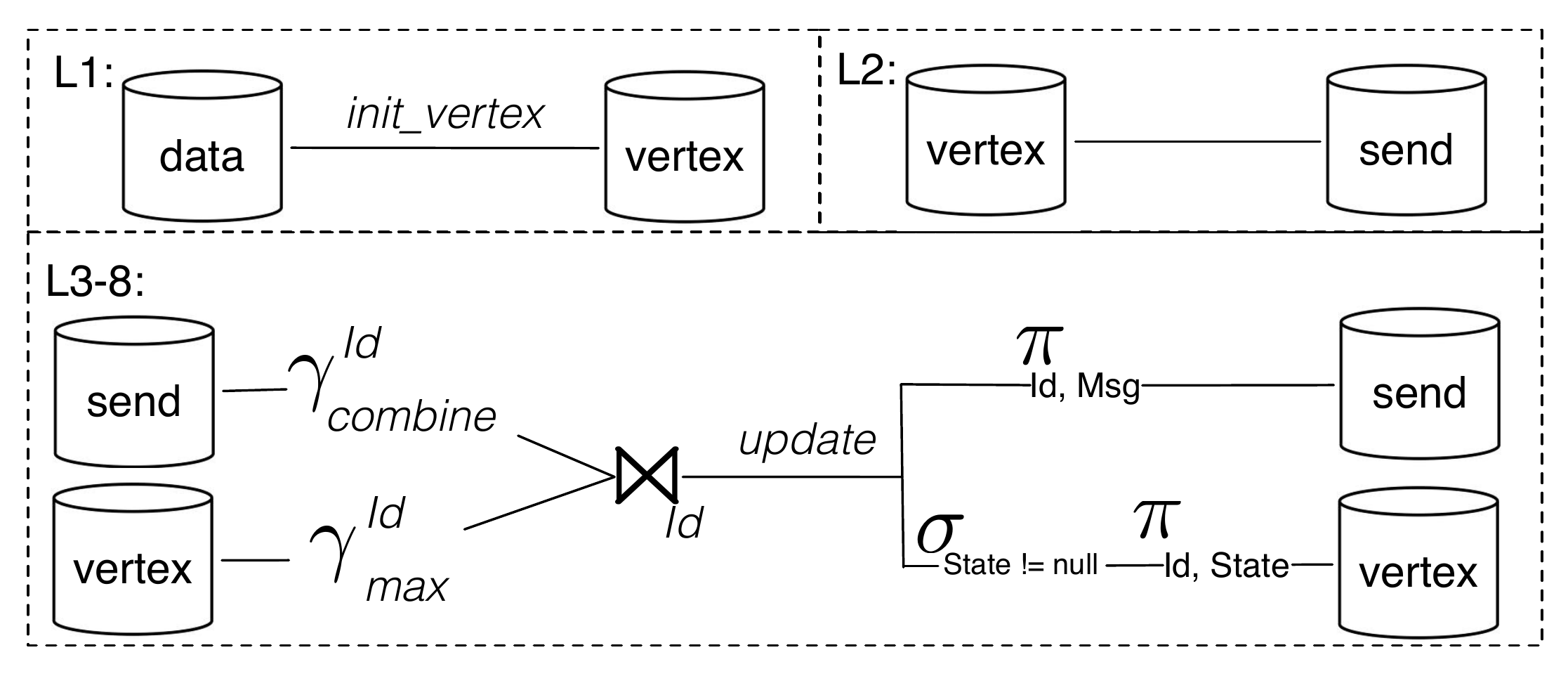}
    \caption{\label{fig:pregel-logical-plan}Logical query plan for Pregel.}
\end{figure}

We can use the same methodology to analyze the Datalog program in
Listing~\ref{lst:datalog-local}.  Here we provide only a brief
summary, since the details are more involved.  XY-stratification
prescribes that the output of the program is computed from an
initialization step with rules \emph{L1} and \emph{L2}, followed by
several iterations where each iteration fires rules in the order
\emph{L3}, $\dots$, \emph{L8}.  Figure~\ref{fig:pregel-logical-plan}
shows the corresponding logical plan in relational algebra.  Data
flows $L1$ and $L2$ initialize the computation, as follows.  In $L1$,
each tuple from the training data is passed to the \ol{init\_vertex}
UDF before being added to the \ol{vertex} dataset.  This triggers $L2$
to generate an initial \ol{send} fact that is destined for each
initial \ol{vertex}.

The dataflow $L3$-$L8$ encodes a single Pregel superstep.  The
\ol{send} dataset is first grouped by the destination vertex
identifier, and each such group of messages is aggregated by the
\ol{combine} UDF\@.  The \ol{vertex} dataset is also grouped by the
vertex identifier, and its most recent state is selected by the
\ol{max} aggregate.  The two results form the \ol{collect} and
\ol{local} IDB predicates (rules \emph{L3}, \emph{L4}, and \emph{L5}),
which are joined along the vertex identifier attribute to produce the
set of vertices with outstanding messages.  The join result is passed
to the \ol{update} function to produce the \ol{superstep} view, which
is subsequently projected along two paths (rule \emph{L6}).  The
bottom path checks for a non-null state object before projecting any
resulting state objects onto the \ol{vertex} dataset (rule \emph{L7}).
The top path projects the set of messages for the next superstep onto
the \ol{send} dataset (rule \emph{L8}).

Overall, it is straightforward to verify that the logical plans match
precisely the logic of the two programming models, which in turn
proves our claim that the Datalog programs are correct.  An equally
important fact is that the logical plan captures the entirety of the
computation, from loading the training data in an initial model to
refining the model through several iterations, along with the
structure of the underlying data flow.  As we discuss in the next
section, this holistic representation is key for the derivation of an
efficient execution plan for the machine learning task.



\section{Physical Dataflow}\label{sec:physical}

In this Section, we present the physical parallel dataflow plans that execute
the Pregel and Iterative Map-Reduce-Update programming models.  We choose the
Hyracks data-parallel runtime~\cite{Borkar:2011ly} as the target platform to
develop and execute these physical plans.  We first give a very brief overview
of Hyracks (Section~\ref{sec:runtime_overview}), then describe the physical
plans for Pregel (Section~\ref{sec:runtime_pregel}) and Iterative
Map-Reduce-Update (Section~\ref{sec:runtime_imr}).  The physical plans
illustrated in this section are used to produce the experimental results
presented in Section~\ref{sec:experiments}.

\subsection{Hyracks Overview}\label{sec:runtime_overview}

Hyracks is a data-parallel runtime in the same general space as
Hadoop~\cite{hadoop} and Dryad~\cite{Isard:2007kx}.  Jobs are submitted to
Hyracks in the form of directed acyclic graphs that are made up of
\emph{operators} and \emph{connectors}.  Operators are responsible for
consuming input partitions and producing output partitions.  Connectors perform
redistribution of data between operators.  Operators are characterized by an
algorithm (e.g., filter, index-join, hash group-by) and input/output data
properties (e.g., ordered-by or partitioned-by some attribute).  Connectors are
classified by a connecting topology and algorithm (e.g., one-to-one connector,
aggregate connector, \text{m-to-n} hash partitioning connector, or
\text{m-to-n} hash partitioning merging connector) as well as by a
materialization policy (e.g., fully pipelining, blocking, sender-side or
receiver-side materializing).

\subsection{Pregel}\label{sec:runtime_pregel}
\eat{
The physical plan for Pregel is a recursive dataflow that executes a series of
Hyracks jobs, each of which makes an iteration and corresponds to a single
Pregel superstep.  A driver program initiates iterations one after the other,
terminating the computation once no more messages are sent from one iteration
to the next.
}

\begin{figure}[!t]
  \centering
    \includegraphics[width=\linewidth]{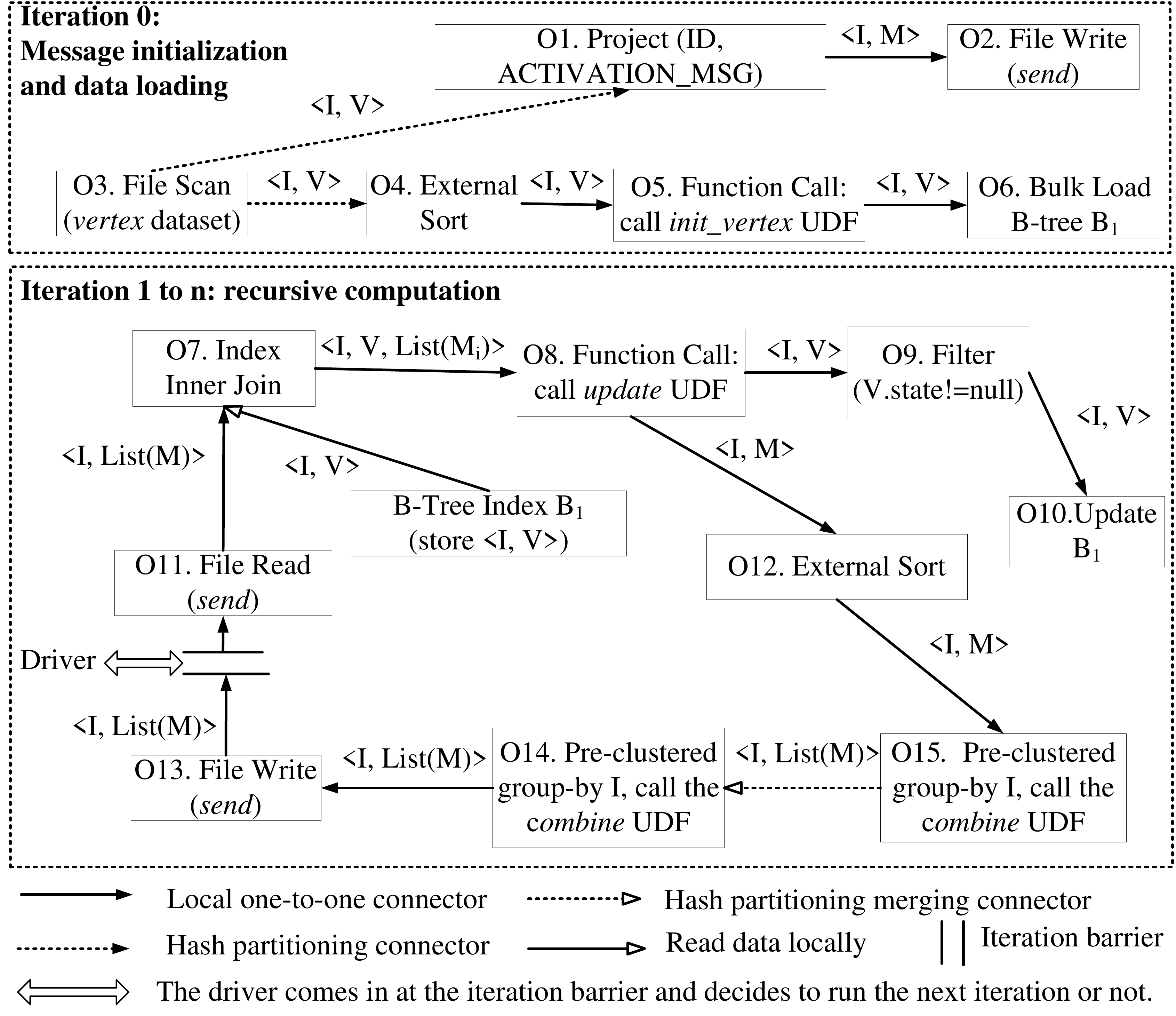}
    \caption{The physical plan for Pregel (operators are labeled \ol{On},
    where \ol{n} is a number).  A connector indicates a type (identified by a
    line and arrow) and a labeled data schema (I: vertex ID; V: vertex data object;
    M: message.)}\label{fig:pregel-physical-plan}
\end{figure}

Figure~\ref{fig:pregel-physical-plan} shows the optimized physical plan for
Pregel.  The top dataflow executes iteration~$0$ and is derived from the
logical plans $L1$ and $L2$ in Figure~\ref{fig:pregel-logical-plan}.  The file
scan operator ($O3$) reads partitions of the input (graph) data, repartitions
each datum by its vertex identifier, and feeds the output to a projection
operator ($O1$) and a sort operator ($O4$).  Operator~$O1$ generates an initial
activation message for each vertex that and writes that result to the \ol{send}
dataset ($O2$).  The sorted tuples from $O4$ are passed to the
\ol{init\_vertex} function, the output of which is then bulk loaded into a
\ol{B-Tree} structure ($O5$).

The bottom dataflow executes iterations until a fixed point: when the \ol{send}
dataset becomes empty (no messages).  This dataflow corresponds to the logical
plan $L3-8$ in Figure~\ref{fig:pregel-logical-plan}.  An iteration starts by
scanning the message lists in the \ol{send} dataset (using operator~$O11$),
which is consumed by an index inner-join operator ($O7$) that ``joins'' with
the vertices in the B-Tree along the identifier attribute.  The join result is
passed to the \ol{update} UDF ($O8$), which produces a new vertex state object
and set of messages.  Operator~$O9$ forwards non-null state objects to update
the B-Tree ($O10$), which occurs locally as indicated by the \text{one-to-one}
connector in the physical plan.  The messages are sorted by operator ($O12$),
and subsequently fed to a pre-clustered group-by operator ($O15$), which groups
messages by the destination vertex ID and uses the \ol{combine} function to
pre-aggregate the messages destined for the same vertex.  A hash partitioning merging
connector shuffles the tuples (using the vertex ID as the shuffle key and a
list of messages as the value) over the network to a consumer (pre-clustered)
group-by operator~$O14$, which again applies the \ol{combine} aggregate
function to the final result before writing to (via $O14$) the new \ol{send}
dataset; occurring on the local machine that also holds the target vertex in the
local B-Tree.  All \ol{send} partitions will report to the driver program,
which determines if another iteration is required: when the \ol{send} dataset
is not empty.

Our translation of the Pregel physical plan from the corresponding logical
plan included a number of physical optimizations:

\noindent\textbf{Early Grouping}: Applies the \ol{combine} function
to the sender-side transfer of the message data in order to reduce the
data volume.

\noindent\textbf{Storage Selection}: In order to efficiently
support primary key updates and avoid the logical \ol{max} aggregation
in Figure~\ref{fig:pregel-logical-plan}, a B-Tree index was chosen
over raw files.

\noindent\textbf{Join Algorithm Selection}: Inputs at
operator~$O7$ are sorted by the join key; allowing an efficient
(ordered) probing strategy.

\noindent \textbf{Order Property}: We selected an order-based group-by strategy
at operator~$O14$ since the input is already sorted.

\noindent \textbf{Shared Scan}: Operators~$O1$ and $O4$ share one file scan.

\eat{ In Section~\ref{sec:experiments}, we evaluate an alternative to
  the plan to Figure~\ref{fig:pregel-physical-plan} that uses a hash
  partitioning connector instead of the hash partitioning merge
  connector between operators~$O15$ and $O14$.  This change requires
  an explicit sort operator be added to the output of operator~$O14$.
  This alternative trades CPU cycles (used to sort) for reduced
  contention (we need to maintain this sorted order) on input streams
  at the repartitioning receiver side.  }

\subsection{Iterative Map-Reduce-Update}\label{sec:runtime_imr}

\begin{figure}
\begin{center}
\includegraphics[width=\linewidth]{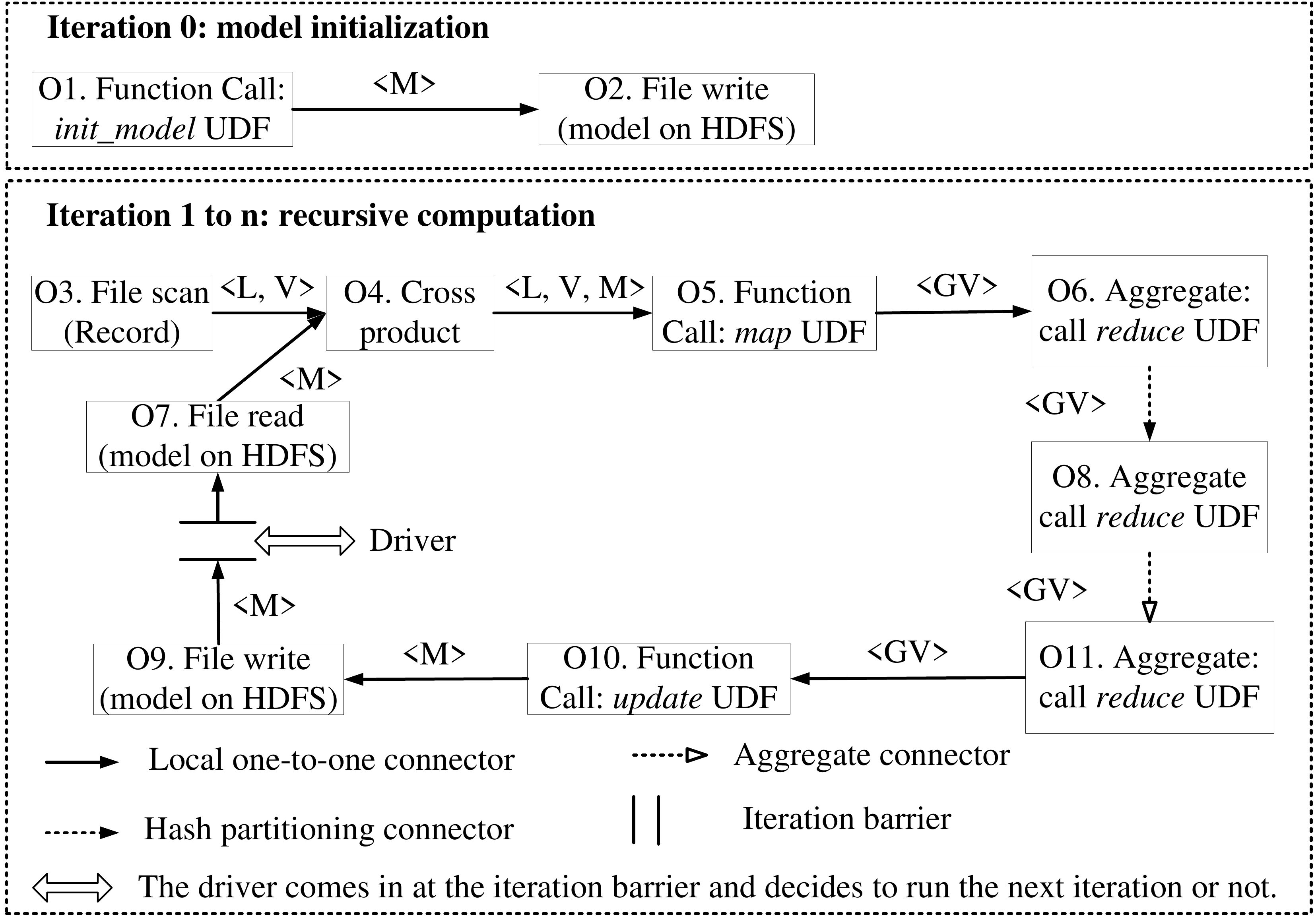}
\caption{The physical plan for Iterative Map-Reduce-Update 
(operators are labeled with \ol{On}, where \ol{n} is a number).
    A connector indicates a type (identified by a line and arrow)
    and a labeled data schema (L: classification label; 
    M: model vector; V: feature vector; GV: (\ol{gradient, loss}) vector.)} 
    \label{fig:bgd-physical-plan}
\end{center}
\end{figure}

We now describe the construction and optimization of a physical plan for the
Iterative Map-Reduce-Update programming model, which is tuned to run Batch
Gradient Descent (BGD).  Figure~\ref{fig:bgd-physical-plan} describes a
physical plan produced by translating the logical query plan in
Figure~\ref{fig:bgd-query-plan} to a physical dataflow.  Iteration~$0$ executes
in the top dataflow, and corresponds to the logical plan $G1$.  Here, we simply
write the initial model to HDFS in operator~$O2$.  The bottom dataflow executes
subsequent iterations until the driver detects termination.  This dataflow
corresponds to the logical plans $G2$--$3$.  At the start of each iteration,
the current model is read from HDFS and paired with the record dataset ($O7$,
$O3$ and $O4$).  A \ol{map} function call operator ($O5$) is passed each record
and the model, and produces a (\ol{gradient, loss}) vector.  The output vectors
are then aggregated through a series of operators ($O6$, $O8$, $O11$) that
apply the \ol{reduce} UDF\@, which in our experiments is a \ol{sum}.  The final
aggregate result is passed to the \ol{update} function call operator ($O10$),
along with the existing model, to produce the next model.  The new model is
written to HDFS ($O7$), where it is read by the driver to determine if more
iterations should be performed.

Two important physical optimization rules were considered when
translating the logical plan into the physical plan:

\noindent\textbf{Early aggregation}: For commutative and associative \ol{reduce}
UDFs (like \ol{sum}), the aggregation should be performed map-local to
reduce the shuffled data volume; thus, $O6$ is included in the plan.

\noindent\textbf{Model volume property}: Large objects
(e.g., vectors in BGD) may saturate a single aggregator's network
connection, resulting in poor performance.  In this case, a layered
aggregation tree must be used to improved performance.  Therefore,
$O8$ is included in the plan.

\section{Experiments}\label{sec:experiments}
In this section, we present experiments comparing the Datalog-derived physical
plans of Section~\ref{sec:physical} to implementations of the same tasks on two
alternative systems: Spark~\cite{spark} for BGD and Hadoop~\cite{hadoop} for
Pregel.  The purpose of these experiments is to demonstrate that a declarative
approach, in addition to shielding ML programmers from physical details, can
provide performance and scalability competitive with current ``best of breed''
approaches.

All experiments reported here were conducted on a 6-rack, 180-machine
Yahoo!\@ Research Cluster.  Each machine has 2 quad-core Intel Xeon
E5420 processors, 16GB RAM, 1Gbps network interface card, and four
750GB drives configured as a JBOD, and runs RHEL 5.6.  The machines
are connected to a top rack Cisco 4948E switch.  The connectivity
between any pair of nodes in the cluster is 1Gbps.  We discuss
system-specific configuration parameters in the relevant subsections.
In Section~\ref{sec:bgd-expr}, we compare our approach against a Spark
implementation on a Batch Gradient Descent task encoded in the
Iterative Map-Reduce-Update programming
model. Section~\ref{sec:pregel-pagerank-expr} presents a PageRank
experiment that runs on a full snapshot of the World-Wide Web from
$2002$ and compares our approach to an implementation based on Hadoop.

\subsection{Batch Gradient Descent}\label{sec:bgd-expr}


\pgfplotscreateplotcyclelist{color and linestyles}{
    black,solid,every mark/.append style={fill=black},mark=square\\%
    black,dashed,every mark/.append style={fill=black},mark=*\\%
    black,densely dashed,mark=star\\%
}

\pgfplotsset{
    width=\columnwidth,
    cycle list name=color and linestyles,
}

\def \bgdSpeedUpCostYMax {1300}
\def \bgdSpeedUpTimeYMax {200}
\def \bgdSpeedUpYMin {0}
\def \bgdSpeedUpXMin {0}
\def \bgdSpeedUpXMax {6}
\def \bgdSpeedUpBarWidth {5}

\begin{figure}[!t]
\centering
\begin{tabular}{c}
\hspace{-13ex}
\begin{minipage}{0.75\columnwidth}
        \begin{tikzpicture}[
    ]
    \begin{axis}[
        axis y line*=left,
        axis x line=none,
        style=black!75!black,
        ybar,
        ylabel=Iteration time (seconds),
        bar width=\bgdSpeedUpBarWidth,
        bar shift=-2.5,
        xmin=\bgdSpeedUpXMin,
        xmax=\bgdSpeedUpXMax,
        ymax=\bgdSpeedUpTimeYMax,
        ymin=\bgdSpeedUpYMin,
    ]
        \addplot+[black] coordinates
        {
            (1, 37.946)
            (2, 15.806)
            (3, 15.407)
            (4, 17.810)
            (5, 20.683)
        };
    \end{axis}
    \begin{axis}[
        axis y line=none,
        xtick={1,...,6}, 
        xticklabels={25, 30, 40, 50, 60},
        xmin=\bgdSpeedUpXMin,
        xmax=\bgdSpeedUpXMax,
        xlabel=Number of machines,
        title=Spark,
        bar width=0,
        legend columns = -1,
        legend to name=bgdspeeduplegend,
        legend entries = {
            Iteration Time;,
            Cost
        },
        ybar,
    ]
        \addplot+[black] coordinates
        {
            (6, 0)
        };
        \addplot+[gray] coordinates
        {
            (6, 0)
        };
    \end{axis}
    \begin{axis}[
        axis y line*=right,
        axis x line=none,
        ybar,
        ylabel=Cost (machine-seconds),
        right,
        style=gray!75!black,
        bar width=\bgdSpeedUpBarWidth,
        bar shift=3.5,
        ylabel style={xshift=-50pt, yshift=-7.2cm},
        xmin=\bgdSpeedUpXMin,
        xmax=\bgdSpeedUpXMax,
        ymax=\bgdSpeedUpCostYMax,
        ymin=\bgdSpeedUpYMin,
    ]
        \addplot+[gray] coordinates
        {
            (1, 948.65)
            (2, 474.18)
            (3, 616.28)
            (4, 890.5)
            (5, 1240.98)
        };
    \end{axis}
    \end{tikzpicture}
\end{minipage}
\\
\hspace{-13ex}
\begin{minipage}{0.75\columnwidth}
        \begin{tikzpicture}[
    ]
    \begin{axis}[
        axis y line*=left,
        axis x line=none,
        style=black!75!black,
        ybar,
        ylabel=Iteration time (seconds),
        bar width=\bgdSpeedUpBarWidth,
        bar shift=-2.5,
        xmax=10,
        ymax=\bgdSpeedUpTimeYMax,
        ymin=\bgdSpeedUpYMin,
    ]
        \addplot+[black] coordinates
        {
            (1, 185.023) 
            (2, 70.4068666666667)
            (3, 46.8957333333333)
            (4, 35.8800666666667)
            (5, 29.142) 
            (6, 25.638) 
            (7, 20.016) 
            (8, 18.195)
            (9, 15.120)
        };
    \end{axis}
    \begin{axis}[
        axis y line=none,
        xtick={1,...,11}, 
        xticklabels={5, 10, 15, 20, 25, 30, 40, 50, 60},
        xmin=0,
        xmax=10,
        xlabel=Number of machines,
        title=Hyracks,
    ]
        \addplot[] coordinates
        {
            (9, 925.115)
        };
    \end{axis}
    \begin{axis}[
        axis y line*=right,
        axis x line=none,
        ybar,
        ylabel=Cost (machine-seconds),
        right,
        style=gray!75!black,
        bar width=\bgdSpeedUpBarWidth,
        bar shift=3.5,
        ylabel style={xshift=-50pt, yshift=-7.2cm},
        xmax=10,
        ymax=\bgdSpeedUpCostYMax,
        ymin=\bgdSpeedUpYMin,
    ]
        \addplot+[gray] coordinates
        {
            (1, 925.115)
            (2, 704.068)
            (3, 703.43595)
            (4, 717.6012)
            (5, 728.55)
            (6, 769.14)
            (7, 800.64)
            (8, 909.75)
            (9, 907.2)
        };
    \end{axis}
    \end{tikzpicture}
\end{minipage}
\end{tabular}

\pgfplotslegendfromname{bgdspeeduplegend}
\caption{\label{fig:bgd-speed-up-graphs}BGD speed-up of Hyracks 
and Spark on Yahoo! News dataset}

\end{figure}
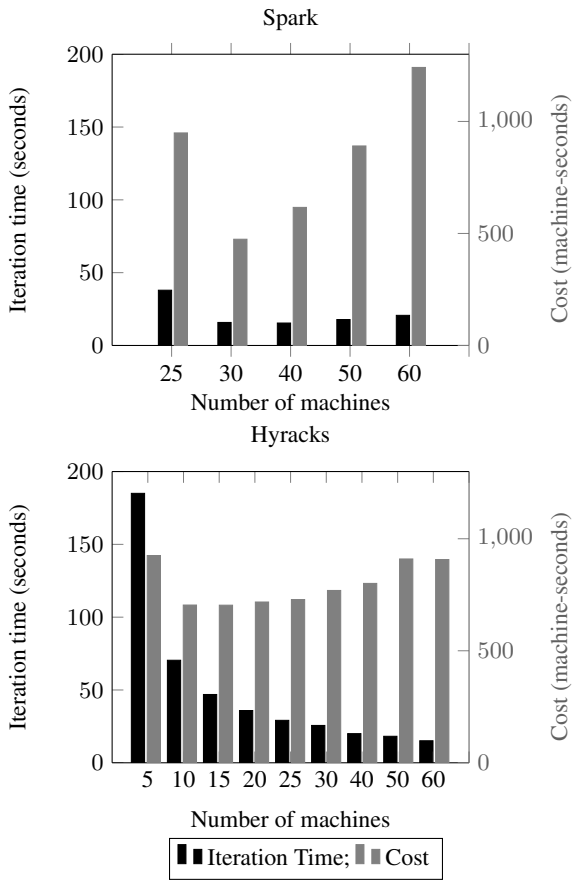

We begin with a Batch Gradient Descent (BGD) task on a real-world dataset drawn
from the web content recommendation domain.  The data consists of $16,557,921$
records sampled from Yahoo!\@ News.  Each record consists of a feature vector
that describes a \ol{(user,content)} pair and a label that indicates whether
the user consumed the content.  The goal of the ML task is to learn a linear
model that predicts the likelihood of consumption for a yet unseen
\ol{(user,content)} pair.  The total number of features in the dataset is
$8,368,084,005$, and each feature vector is sparse: users are only interested
in a small subset of the content.  The dataset is stored in HDFS and, before
running the job, it is perfectly balanced across all machines used in the
experiment.  That is, each machine is assigned an equal number of records.

We report results for running this task in Hyracks (using the physical plan in
Figure~\ref{fig:bgd-physical-plan}) to Spark.  The Spark code was organized
similarly and verified by the system author (Matei~Zaharia).  Specifically, in
the first step, we read each partition from HDFS and convert it to an internal
record: sparse vectors are optimized in a compact form.  In Hyracks, each
partition is converted into a binary file representation and stored in the
local file system.  For Spark, we make an explicit cache call that ``pins'' the
records in memory.  Both systems then execute a fixed number of iterations,
each of which executes a single \ol{map}, \ol{reduce}, and \ol{update} step.

In the \ol{map} step, we make a pass over all the records in a given internal
partition and compute a single \ol{(gradient, loss)} value based on the current
model, which resides in HDFS\@.  The \ol{reduce} step sums all the
\ol{(gradient, loss)} values produced by individual \ol{map} tasks to a single
aggregate value.  We use pre-aggregators in both systems to optimize the
computation of these sums.  In Spark, we use a single layer of $\sqrt{\text{num
partitions}}$ pre-aggregators.  Hyracks performs local pre-aggregation on each
machine (holding four \ol{map} partitions) followed by a single layer of
$\sqrt{\text{num map machines}}$ pre-aggregators.  We also evaluate an
alternative (more optimal) Hyracks configuration, which again performs a local
pre-aggregation but then uses a 4-ary aggregation tree (a variable-height
aggregation tree where each aggregator receives at most 4 inputs).  The Spark
API did not allow us to use a (\ol{map}) machine local pre-aggregation
strategy, and there is no system support for such optimizations.  The final
\ol{update} step takes the aggregated result and the current model and produces
a new model that is written to HDFS~\footnote{Spark exposes this operation
through a ``broadcast'' variable.} for use by the next iteration's \ol{map}
step.

We now present two sets of experiments.  First, we identify the
cost-optimal number of machines that each system should use to process
a fixed-size dataset.  Using the cost-optimal configuration, we
measure the scalability of each system by proportionately increasing
the dataset size and number of machines used.

\subsubsection{Cost-optimal configuration for fixed-size data}

The goal of this experiment is to determine the optimal number of
machines that should be used to perform the BGD task for a fixed-size
dataset on Spark and Hyracks.  We measure cost in terms of
machine-seconds ($number\ of\ machines \times average\ iteration\
time$) and look for a cluster size that minimizes it.
Figure~\ref{fig:bgd-speed-up-graphs} reports both time and cost,
averaged over five iterations, as we increase the number of machines
while keeping the total dataset size fixed at $\sim$80GB\@.  Increasing
the number of machines generally improves the iteration time, but
diminishing returns due to increasing overhead make it
cost-inefficient.  From this experiment, we identify the cost optimal
configurations to be $30$ machines for Spark and $10$ machines (giving
preference to fewer machines) for Hyracks.  Note that Hyracks could
use an arbitrarily small number of machines since it supports
out-of-core computations.
Spark, however, is restricted to main-memory, and as a result,
requires at least $25$ machines to run this experiment.


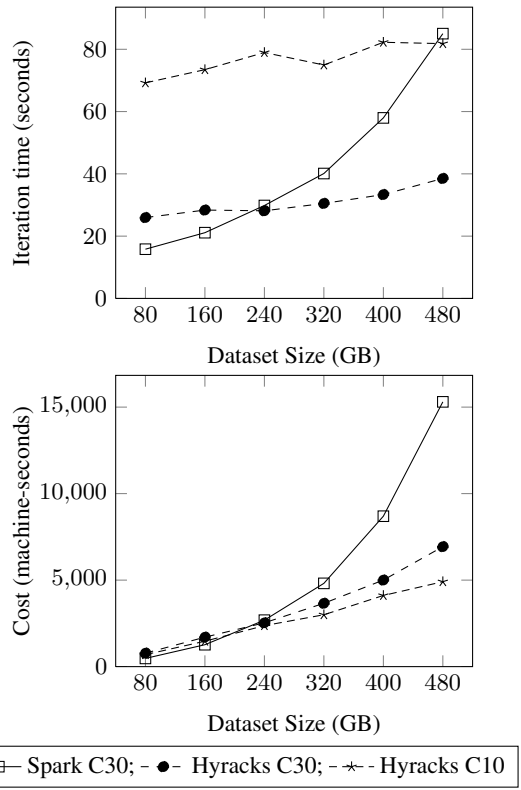
\begin{figure}[!t]
\centering
\begin{tabular}{c}
\begin{minipage}{0.75\columnwidth}
    \begin{tikzpicture}[
    ]
    \begin{axis}[
        xlabel=Dataset Size (GB),
        ylabel=Iteration time (seconds),
        legend columns = -1,
        legend entries = {
            Spark C30;,
            Hyracks C30;,
            Hyracks C10
        },
        legend to name=bgdlegend,
        xtick=data,
        ymin=0,
    ]
        \addplot+[sharp plot] coordinates
        {
            (80, 15.806)
            (160, 21.106)
            (240, 29.837)
            (320, 40.105)
            (400, 57.964)
            (480, 85.020)
        };

        \addplot+[sharp plot] coordinates
        {
            (80, 25.962)
            (160, 28.370)
            (240, 28.135)
            (320, 30.504)
            (400, 33.350)
            (480, 38.520)
        };

        \addplot+[sharp plot] coordinates
        {
            (80, 69.215)
            (160, 73.458)
            (240, 78.965)
            (320, 74.948)
            (400, 82.248)
            (480, 81.810)
        };
    \end{axis}
    \end{tikzpicture}
    \label{fig:bgd-scale-up-time}
\end{minipage}
\\
\begin{minipage}{0.75\columnwidth}
    \begin{tikzpicture}[
    ]
    \begin{axis}[
        xlabel=Dataset Size (GB),
        ylabel=Cost (machine-seconds),
        scaled ticks = false,
        xtick=data,
        ymin=0,
    ]
        \addplot+[sharp plot] coordinates
        {
            (80, 474.18)
            (160, 1266.36)
            (240, 2685.33)
            (320, 4812.6)
            (400, 8694.6)
            (480, 15303.6)
        };

        \addplot+[sharp plot] coordinates
        {
            (80, 778.86)
            (160, 1702.2)
            (240, 2532.15)
            (320, 3660.48)
            (400, 5002.5)
            (480, 6933.6)
        };

        \addplot+[sharp plot] coordinates
        {
            (80, 692.15)
            (160, 1469.16)
            (240, 2368.95)
            (320, 2997.92)
            (400, 4112.4)
            (480, 4908.6)
        };
    \end{axis}
    \end{tikzpicture}
    \label{fig:bgd-scale-up-time}
\end{minipage}
\end{tabular}
\pgfplotslegendfromname{bgdlegend}
\caption{\label{fig:bgd-scale-up-graphs}BGD scale-up of Hyracks vs.\@ Spark}

\end{figure}

\subsubsection{Scalability}

Given the cost-optimal configuration, we now explore the scalability of each
system as we proportionally increase the input training data size and number of
machines.  To scale up the data, we duplicated and randomly shuffled the
original data.  The cost-optimal settings are captured by the following two
cluster-size-for-data-size configurations:
%
\begin{description}{\leftmargin=0em} \itemsep 0pt \itemindent -2em
    \listparindent 0pt \parskip 0pt \item[C10]: 10 nodes per 80GB (Hyracks
    cost-optimal) \item[C30]: 30 nodes per 80GB (Spark cost-optimal)
\end{description}
We executed both Hyracks and Spark on configuration \text{C30}, but we only ran
Hyracks on \text{C10} since Spark was unable to retain this much data in the
given amount of main memory.  Figure~\ref{fig:bgd-scale-up-graphs} reports the
results of this scalability experiment, showing iteration time at the top and
cost at the bottom.  The x-axis for both graphs range over increasing data
sizes.  Each configuration adds the baseline number of nodes to match the data
size.  Example: At data size $160$GB, configuration \text{C10} uses $20$
machines and \text{C30} uses $60$ machines.

As we scale up, we expect the \ol{map} part of the iteration to scale
perfectly.  However, as we add more partitions, we create more
intermediate results that need to be transferred over the network to
the \ol{reduce} aggregation.  It turns out that the amount of network
traffic between the \ol{map} nodes and the intermediate
pre-aggregators is linear in the number of \ol{map} nodes, and the
work done in reducing the intermediate results is proportional to the
square root of the \ol{map} nodes.  Thus, we expect a growth in
completion time as we scale up the number of \ol{map} nodes.  We
clearly see this trend for the execution time of Spark.  However, the
growth in completion time for Hyracks is much slower, benefiting from
the machine-local aggregation strategy.  Hyracks also uses a
packet-level fragmentation mechanism that achieves better overlap in
the network transfer and aggregation of intermediate results;
receiving aggregators can immediately start reducing each fragment
independently while other fragments are in transit.  Spark on the
other hand, waits for the complete \ol{(gradient, loss)} result---a
$\sim$16MB size vector---to be received before incorporating it into
the running aggregate. Additionally, Spark faces other system-level
bottlenecks due to its use of a stock data-transfer library to move
data between processes, while Hyracks has an efficient custom
networking layer built for low-latency high-throughput data transfer.

For data sizes $80$GB and $160$GB, Spark finishes slightly earlier than
Hyracks.  There are two factors that contribute to this phenomenon.  The first
is that Hyracks currently has slightly higher overhead in the \ol{map} phase in
how it manages the input data.  As mentioned earlier, Hyracks uses the local
file system to store a binary form of the data and relies on the file system
cache to avoid disk I/O.  Accessing data through the file system interface on
each iteration, and the data-copies across this interface, account for slightly
larger map times.  Spark, on the other hand, loads the input data into the JVM
heap and no data copies are necessary during each iteration.  A second cause of
slow down for Hyracks is that the local aggregation step in Hyracks adds
latency, but does not help much in lowering the completion time of an iteration
for the $80$GB case.  In the $160$GB case, the benefit of local aggregation is
still out-weighed by the latency introduced.  In our experimental setup, each
rack has $30$ machines resulting in rack-local computation in the $80$GB case.
In the $160$GB case, the computation is spread across two racks.  Since enough
bandwidth is available within a rack, the local aggregation does not appear to
pay off in these two cases.  Our planned solution to the first problem is to
take on more of the buffer-management in Hyracks to reduce or eliminate data
copies for the data that resides in memory, but still use the file system so
that we can scale to data larger than main memory.  The second problem
motivates the need for a runtime optimizer that decides when it is appropriate
to use local combiners to solve the problem for a given physical cluster.  In
the future, the optimizer that generates the Hyracks job for the Batch Gradient
Descent task will be expected to make the correct choice with regards to local
aggregation.

As we scale up the data beyond $160$GB, we see that the Hyracks system
shows better scale up characteristics for the reasons mentioned above.
The cost curve in the bottom graph shows a similar trend to that of
the time curve.  As we linearly increase the data and cluster size,
the cost to solve the problem with Spark grows much faster than the
cost to solve the problem with Hyracks.

\subsection{PageRank}\label{sec:pregel-pagerank-expr}

The goal of PageRank is to rank all web pages by their relative importance.  We
perform our experiments on Yahoo!'s publicly available webmap
dataset~\cite{webmap}, which is a snapshot of the World-Wide Web from $2002$.
The data consists of $1,413,511,393$ vertices, each of which represents a
web-page.  Each record consists of a source vertex identifier and an array of
destination vertex identifiers forming the links between different webpages in
the graph.  The size of the decompressed data is $70$GB, which is stored in
HDFS and evenly balanced across all participating machines.

We compare the performance of Hyracks (using the physical plan in
Figure~\ref{fig:pregel-physical-plan}) to an implementation of
PageRank in Hadoop.  The Hadoop code for PageRank consists of a job
that first joins the ranks with the corresponding vertices.  This is
followed by a grouping job that combines contribution rank values from
``neighboring'' vertices to compute the new rank.  Hyracks executes
the whole iteration of the PageRank algorithm in a single job.  For
both systems we perform $10$ iterations.

We follow the same methodology as above: First, we identify the
cost-optimal number of machines for each system using a fixed-size
dataset ($70$GB).  Then we explore the scalability of Hyracks and
Hadoop by running PageRank against proportionately increasing dataset
sizes and number of machines, using the cost-optimal machine
configurations.


\pgfplotscreateplotcyclelist{color and linestyles}{
    black,dashed,every mark/.append style={fill=black},mark=*\\%
    black,densely dashed,mark=star\\%
    black,solid,every mark/.append style={fill=black},mark=square\\%
}

\pgfplotsset{
    width=\columnwidth,
    cycle list name=color and linestyles,
}


\def \hyracksSpeedUpYMin {0}
\def \hyracksSpeedUpBarWidth {5}

\begin{figure}[!t]
\centering
\begin{tabular}{c}
\hspace{-13ex}
\begin{minipage}{0.75\columnwidth}
        \begin{tikzpicture}[
    ]
    \begin{axis}[
        axis y line*=left,
        style=black!75!black,
        ybar,
        ylabel=Iteration time (seconds),
        bar width=\bgdSpeedUpBarWidth,
        bar shift=-2.5,
        xtick=data,
        ymin=\bgdSpeedUpYMin,
        scaled ticks = false,
        title=Hadoop,
        xlabel=Number of machines
    ]
        \addplot+[black] coordinates
        {
            (31, 2128.549)
            (60, 1184.116)
            (88, 701.411)
            (117, 550.340)
            (146, 472.152)
        };
    \end{axis}
    \begin{axis}[
        axis y line=none,
        axis x line=none,
        bar width=0,
        legend columns = -1,
        legend to name=pregelspeeduplegend,
        legend entries = {
            Iteration Time;,
            Cost
        },
        ybar,
    ]
        \addplot+[black] coordinates
        {
            (6, 0)
        };
        \addplot+[gray] coordinates
        {
            (6, 0)
        };
    \end{axis}
    \begin{axis}[
        axis y line*=right,
        axis x line=none,
        ybar,
        ylabel=Cost (machine-seconds),
        right,
        style=gray!75!black,
        bar width=\bgdSpeedUpBarWidth,
        bar shift=3.5,
        ylabel style={xshift=-50pt, yshift=-7.2cm},
        scaled ticks = false,
        ymin=\bgdSpeedUpYMin
    ]
        \addplot+[gray] coordinates
        {
            (31, 65985.010)
            (60, 71046.936)
            (88, 61724.153)
            (117, 64389.831)
            (146, 68934.173)
        };
    \end{axis}
    \end{tikzpicture}
\end{minipage}
\\
\hspace{-13ex}
\begin{minipage}{0.75\columnwidth}
        \begin{tikzpicture}[
    ]
    \begin{axis}[
        axis y line*=left,
        style=black!75!black,
        ybar,
        ylabel=Iteration time (seconds),
        bar width=\bgdSpeedUpBarWidth,
        bar shift=-2.5,
        ymin=\bgdSpeedUpYMin,
        xtick=data,
        xlabel=Number of machines,
        title=Hyracks
    ]
        \addplot+[black] coordinates
        {
            (31, 186.137)
            (60, 98.500)
            (88, 67.993) 
            (117, 55.091) 
            (146, 49.763) 
        };
    \end{axis}
    \begin{axis}[
        axis y line*=right,
        axis x line=none,
        ybar,
        ylabel=Cost (machine-seconds),
        right,
        style=gray!75!black,
        bar width=\bgdSpeedUpBarWidth,
        bar shift=3.5,
        ylabel style={xshift=-50pt, yshift=-7.2cm},
        ymin=\bgdSpeedUpYMin,
    ]
        \addplot+[gray] coordinates
        {
            (31, 5770.240)
            (60, 5910.016)
            (88, 5983.394)
            (117, 6445.656)
            (146, 7265.355)
        };
    \end{axis}
    \end{tikzpicture}
\end{minipage}
\end{tabular}

\pgfplotslegendfromname{pregelspeeduplegend}

\caption{\label{fig:pregel-speed-up-graphs}PageRank speed-up of Hyracks vs. Hadoop}
\end{figure}
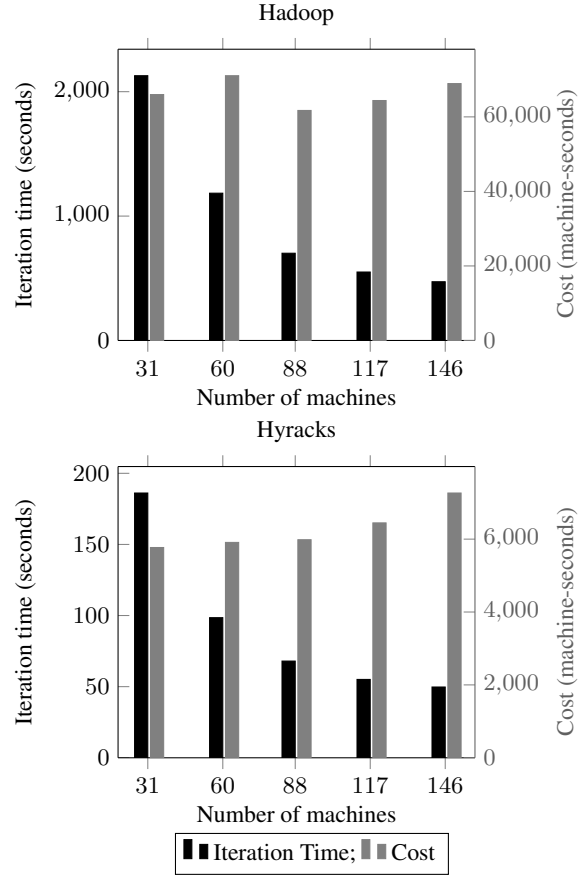

\subsubsection{Cost-optimal configuration for fixed-size data}

In this experiment, we determine the cost-optimal number of machines to be used
for a fixed-size ($70$GB) dataset on Hadoop and Hyracks.
Figure~\ref{fig:pregel-speed-up-graphs} reports the average iteration time and
the cost in terms of machine-seconds ($number\ of\ machines \times average\
iteration\ time$) for different number of machines.  The iteration time in both
systems is decreases as we add more machines.  Hadoop's iteration cost
fluctuates as we increase the number of machines, whereas Hyracks' cost
increases slowly.  Also, we note the following effects: 1) As we add more
machines, the benefit obtained from local combiners gradually diminishes, and
2) the repartitioning step becomes the bottleneck in Hadoop.  Hadoop's
implementation of PageRank needs to shuffle both the graph data (which is
invariant across iterations) and the rank contributions, leading to far more
data movement over the network than the PageRank plan in Hyracks.  Hyracks
moves around (shuffles) only the rank contributions over the network, while
caching the loop-invariant graph data at the same nodes across iterations.
This extra data movement accounts for most of the order-of-magnitude increase
in iteration time experienced by Hadoop when comparing to Hyracks.

The cost-optimal configuration is $31$ machines for Hyracks and $88$
machines for Hadoop per $70$GB of data as per
Figure~\ref{fig:pregel-speed-up-graphs}.

\subsubsection{Scalability}

To scale up the data size, we duplicated the original graph data and renumbered
the duplicate vertices by adding each identifier with the largest vertex
identifier in the original graph.  Thus, one duplication creates a graph that
has twice as many vertices in two disconnected subgraphs.  The nodes in the
resulting graphs were randomly shuffled before loading the data onto the
cluster.  While we recognize that this does not follow the structure of the
web, this experiment is concerned with the behavior of the dataflow rather than
the actual result of the PageRank algorithm.

Based on the cost-optimal results from the speed-up we derive the following
two configurations:
\begin{description}{\leftmargin=0em}
    \itemsep 0pt \itemindent -2em \listparindent 0pt \parskip 0pt
    \item[C31]: 31 machines per $70$GB (Hyracks cost-optimal)
    \item[C88]: 88 machines per $70$GB (Hadoop cost-optimal)
\end{description}

\pgfplotscreateplotcyclelist{color and linestyles}{
    black,solid,every mark/.append style={fill=black},mark=*\\%
    black,densely dashed,mark=square\\%
}

\pgfplotsset{
    width=\columnwidth,
    cycle list name=color and linestyles,
}

Table~\ref{tab:pregel-scalup} shows that Hyracks PageRank performance
for data sizes $70$GB and $140$GB is an order-of-magnitude faster and
cheaper than Hadoop (in Hadoop's optimal configuration C88) owing to
more data movement over the network, as described above. Both systems
scale similarly as we grow the graph data and the size of the cluster.

\begin{table}
\small
\begin{tabular}{|r|r|r|r|}
\cline{1-4}
           \textbf{Configuration} & \textbf{Dataset Size(GB)} &  \textbf{Iteration Time(s)} &\textbf{Cost} \\
\hline      Hyracks-C88 &70 & 67.993 & 5983.394\\
\cline{1-4} Hadoop-C88 & 70& 701.411& 61724.153\\
\cline{1-4} Hyracks-C88 & 140& 84.970 &14869.750 \\
\cline{1-4} Hadoop-C88 & 140& 957.727 & 167602.196  \\
\cline{1-4} Hyracks-C31 & 70& 186.137 & 5770.240   \\
\cline{1-4} Hyracks-C31 & 140& 208.444 & 12506.658   \\
\cline{1-4}
\end{tabular}
\caption{PageRank scale-up of Hyracks vs.\@ Hadoop} \label{tab:pregel-scalup}
\end{table}
\subsubsection{Comparing Different Hyracks Plans}

\begin{figure}[!t]
\centering
\begin{tabular}{c}
\begin{minipage}{0.75\columnwidth}
    \begin{tikzpicture}[
    ]
    \begin{axis}[
        xlabel=Dataset Size (GB),
        ylabel=Iteration time (seconds),
        xtick=data,
        legend columns = -1,
        legend entries = {
           Hyracks;,
           Hyracks (hash partitioning connector)
        },
        legend to name=pregellegend,
    ]
        \addplot+[sharp plot] coordinates
        {
            (70, 186.137)
            (140, 208.444)
            (210, 235.317)
            (280, 357.189)
            (350, 385.654)
        };

        \addplot+[sharp plot] coordinates
        {
            (70, 190.148)
            (140, 213.823)
            (210, 239.309)
            (280, 295.673)
            (350, 366.589)
        };
    \end{axis}
    \end{tikzpicture}
    \label{fig:pregel-speed-up-iteration-time}
\end{minipage}
\\
\begin{minipage}{0.75\columnwidth}
    \begin{tikzpicture}[
    ]
    \begin{axis}[
        xlabel=Dataset Size (GB),
        ylabel=Cost (machine-seconds),
        xtick=data,
        scaled ticks = false
    ]
        \addplot+[sharp plot] coordinates
        {
           (70, 5770.240)
           (140, 12506.658)
           (210, 20707.870)
           (280, 41791.165)
           (350, 56305.543)
        };

        \addplot+[sharp plot] coordinates
        {
            (70, 5894.580)
            (140, 12829.404)
            (210, 21059.228)
            (280, 34593.693)
            (350, 53522.053)
        };
    \end{axis}
    \end{tikzpicture}
    \label{fig:bgd-scale-up-time}
\end{minipage}
\end{tabular}
\pgfplotslegendfromname{pregellegend}

\caption{\label{fig:pregel-scale-up-graphs}PageRank scale-up of Hyracks alternative plans}
\end{figure}
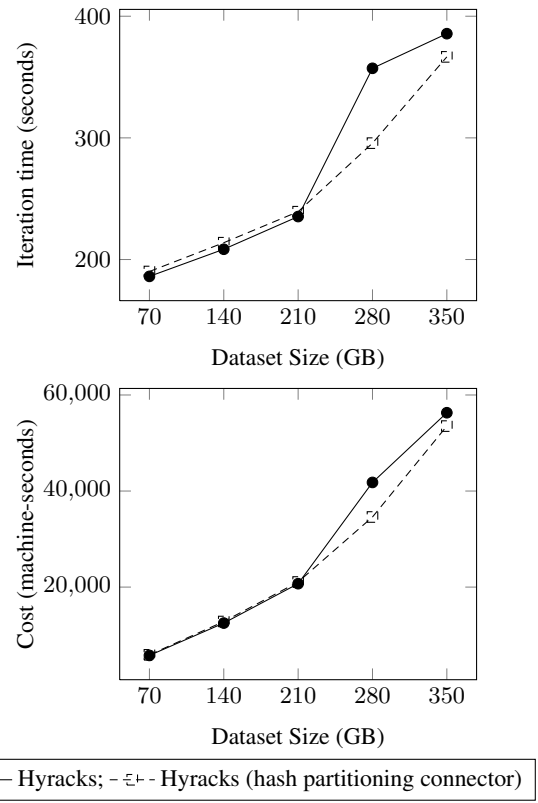

To further investigate performance differences associated with
alternate physical data movement strategies, we tried rerunning
Hyracks with a slight variation in the connector used to redistribute
the messages from the message combiners (O15) to the message reducers
(O14) in the plan shown in Figure~\ref{fig:pregel-physical-plan}.
We replaced the hash partitioning merging connector with a simpler hash
partitioning connector. While the original merging connector maintained
the sorted order of messages as they were received from each combiner,
the hash partitioning connector merges data from any sender in the
order it is received, thus destroying the sorted property. In order to
get the sorted property back, we added an explicit sorter before
feeding the messages into O14.
Figure~\ref{fig:pregel-scale-up-graphs} shows the iteration times and
the cost of iterations of the two Hyracks plans as we scale up the
graph size and the number of machines used to compute the PageRank
using configuration \text{C31}.

We see that for smaller data and cluster sizes ($70$GB to $210$GB),
the Hyracks plan with the hash partitioning merging connector runs
faster than the one with the hash connector with explicit
sorting. This is because the former plan does less work in maintaining
the sort order because the merge exploits the sorted input property to
merge the incoming data at receiver using a priority queue, much like
the merge phase used in external sorting. However, each receiver of
the merge process selectively waits for data to arrive from a specific
sender as dictated by the priority queue. Temporary slowness on behalf
of a sender at a time when a receiver needs data from it leads to a
stall in the merge pipeline. Although other senders are willing to
send data, they have to wait for the specific sender that the receiver
is waiting for to make data available.  The resulting degradation in
the iteration time is observable as the size of the cluster grows to
data sizes of $280$GB and $350$GB\@.  At these sizes, the savings in
work achieved by the hash partitioning merging connector are far
outweighed by the coordination overhead introduced by the merge
process.
This tradeoff is evidence that an optimizer is ultimately essential to
identify the best configuration of the runtime plan to use in order to
solve the Pregel problem.

\subsection{Discussion}\label{sec:expr-discussion}

One might wonder why Hadoop was the chosen for the reference
implementation of the Pregel runtime plan.  Before we compared our
system with Hadoop, we also tried to compare it with three other
``obvious candidate'' systems, namely Giraph~\cite{giraph},
Mahout~\cite{mahout}, and Spark~\cite{spark}.  What we discovered (the
hard way!) is that none of those systems was able to run PageRank for
the Yahoo! webmap dataset, even given all 6 racks (175 machines), due
to design issues related either to memory management (Giraph and
Spark) or to algorithm implementation (Mahout).

An interesting observation regarding the Spark user model was the
process involved in implementing a 1-level aggregation tree.  In order
to perform a pre-aggregation in Spark we had to explicitly write---in
user facing code---an intermediate ``\ol{reduceByKey}'' step that
subsequently feeds the final (global) \ol{reduce} step.  We assigned a
random number (using \ol{java.lang.Random.nextInteger} (modulo the
number of pre-aggregators) as the key to the \ol{(gradient, loss)}
record from the \ol{map} step.  Ideally, such an optimization should
be captured by the system, and not in user code.

\section{Related Work}\label{sec:related}

Our work builds upon and extends prior results
from of a number of different research areas.

{\bf Parallel database systems} such as Gamma~\cite{Dewitt-Gamma},
Teradata~\cite{teradata}, and GRACE~\cite{grace} applied partitioned-parallel
processing to data management, particularly query processing, over two decades
ago.  The introduction of Google's MapReduce system~\cite{Dean:2004uq},
based on similar principles, led to the recent flurry of work
in MapReduce-based data-intensive computing.  Systems like
Dryad~\cite{Isard:2007kx} and Hyracks~\cite{Borkar:2011ly}
have successfully made the case for supporting a richer set
of data operators beyond map and reduce as well as a richer
set of data communication patterns.

{\bf High-level language abstractions} like Pig~\cite{Olston:2008kx},
Hive~\cite{hive}, and DryadLINQ~\cite{Isard:2009vn} reduce the
accidental complexity of programming in a lower-level dataflow
paradigm (e.g., MapReduce).  However, they do not support iteration as
a first class citizen, instead focusing on data processing pipelines
expressible as directed acyclic graphs.  This forces the use of
inefficient external drivers when iterative algorithms are required to
tackle a given problem.

{\bf Iterative extensions to MapReduce} like HaLoop~\cite{haloop} and
PrIter~\cite{priter} were the first to identify and address the need for
runtime looping constructs.  HaLoop uses a ``sticky scheduling'' policy to
place map and reduce tasks in downstream jobs on the same physical machines
with the same inputs.  In Hyracks, the job client is given control over the
task placement, which we use to implement a similar policy.  PrIter uses a
key-value storage layer to manage its intermediate MapReduce state, and it also
exposes user-defined policies that can prioritize certain data to promote fast
algorithmic convergence.  However, those extensions still constrain
computations to ``map'' and ``reduce'' functions, while Hyracks allows more
flexible computations and forms of data redistribution for optimizing machine
learning tasks.

{\bf Domain-specific programming models} like Pregel~\cite{pregel},
GraphLab~\cite{graphlab}, and Spark~\cite{spark}, go beyond one-off
implementations for specific algorithms (e.g.~\cite{Alekh-Agarwal:2011fk,
Weimer:2010fk}), to general purpose systems that capture a specific class of ML
tasks.  Of these, Spark is the most general, but it lacks a runtime optimizer and
support for out-of-core operators, making it hard to tune.  GraphLab and Pregel
expose a graph-oriented programming model and runtime that is very appropriate
for some ML tasks but suboptimal for others.  GraphLab supports asynchronous
execution, which lends itself to graphical model machine learning.

{\bf RDBMS extensions} have been proposed that provide direct support for ML
tasks.  In Tuffy~\cite{tuffy}, Markov Logic Networks are represented as
declarative rules in first-order-logic, and from there, optimized into an
efficient runtime plan by a RDBMS\@.  MadLib~\cite{madlib} maps linear algebra
operations, such as matrix multiplies, to SQL queries that are then compiled
and optimized for a parallel database system.  These approaches are limited
to single pass algorithms (i.e., closed form solutions) or require the use
of an external driver for iterative algorithms.

{\bf Datalog extensions} have also been proposed for implementing ML
tasks.  Atul and Hellerstein~\cite{Atul-thesis} use Overlog---a
distributed Datalog-like declarative language---to elegantly capture
probabilistic inference algorithms; among them, a Junction Tree
Running Intersection Property expressed in a mere seven Overlog rules.
Dyna uses a Datalog extension to capture statistical Artificial
Intelligence algorithms as systems of equations, which relate
intensional and extensional data to form structured prediction
models~\cite{eisner-filardo-2011}.  Dyna compiles such model
specifications into efficient code.

Our approach shares many aspects with various of the aforementioned systems,
yet it is unlike any one of those.  To the best of our knowledge, this
paper has proposed the first distributed, out-of-core-capable runtime for
Datalog aimed at supporting several end-user programming models at once,
thereby unifying machine learning and ETL processes within a single
framework and on a single, and scalable, runtime platform.

\section{Conclusion}\label{sec:conclusion}


The growing demand for machine learning is pushing both industry and academia
to design new types of highly scalable iterative computing systems.  Examples
include Mahout, Pregel, Spark, Twister, HaLoop, and PrItr.  However, today's
specialized machine learning platforms all tend to mix logical representations
and physical implementations.  As a result, today's platforms 1) require
their developers to rebuild critical components and to hardcode optimization
strategies and 2) limit themselves to specific runtime implementations that
usually only (naturally) fit a limited subset of the potential machine learning
workloads.  This leads to the current state of practice, wherein the
implementation of new scalable machine learning algorithms is very
labor-intensive and the overall data processing pipeline involves multiple
disparate tools hooked together with file- and workflow-based glue.

In contrast, we have advocated a declarative foundation on which specialized
machine learning workflows can be easily constructed and readily tuned.  We
verified our approach with Datalog implementations of two popular programming
models from the machine learning domain: Pregel, for graphical algorithms, and
Iterative Map-Reduce-Update, for deriving linear models.  The resulting Datalog
programs are compact, tunable to a specific task (e.g., Batch Gradient Descent
and PageRank), and translated to optimized physical plans.  Our experimental
results show that on a large real-world dataset and machine cluster, our
optimized
plans are very competitive with other systems that target the given class
of ML tasks.  Furthermore, we demonstrated that our approach can offer a
plan tailored to a given target task and data for a given machine resource
allocation.  In contrast, in our large experiments, Spark failed due to
main-memory limitations and Hadoop succeeded but ran an order-of-magnitude
less efficiently.

The work reported here is just a first step.  We are currently developing
the ScalOps query processing components required to automate the remaining
translation steps from Figure~\ref{fig:architecture}; these include the Planner/Optimizer
as well as a more general algebraic foundation based on extending the
Algebricks query algebra and rewrite rule framework of ASTERIX~\cite{ASTERIX}.
We also plan to investigate support for a wider range of machine learning
tasks and for a more asynchronous, GraphLab-inspired programming model
for encoding graphical algorithms.

\bibliographystyle{plain}
\bibliography{paper}

\pagebreak
\begin{appendix}

\section{Batch Gradient Descent}\label{sec:bgd}

Much of supervised machine learning can be cast as a convex optimization
problem.  In supervised machine learning, we are given a database of pairs
$(x,y)$, where $x$ is a data point and $y$ is a label.  The goal is to find a
function $f_w(x)$ that can predict the labels for the yet unseen examples $x$.
Depending on the type of $y$, this definition specializes into many machine
learning tasks: regression, (binary and multi-class) classification, logistic
regression and structured prediction are some examples.

Learning the function $f_w$ amounts to searching the space of parameterized
functions.  The parameters are also called model and are typically referred to
as~$w$.  Hence, the search for $f_w$ is the search for $w$.  This search
problem is guided by a \emph{loss function}~$l(f_w(x), y)$ that measures the
divergence between a prediction~$f_w(x)$ and a known label~$y$.  A large class
of machine learning problems also include a \emph{regularizer}
function~$\Omega(w))$ that measures the complexity of~$f_w$.  Following Occam's
razor---all things equal favor a simpler model---the regularizer is added to
the loss to form the template of a supervised machine learning problem:

\begin{equation}\label{eq:regrisk}
  \hat{w} = \argmin{w} \left(\lambda \Omega(w) + \sum_{(x,y) \in D} l\left(f_w(x),y\right)\right)
\end{equation}

The loss function is sometimes referred to as (empirical) risk, and therefore
the above optimization problem is known as \emph{regularized risk minimization}
in the literature.  From a dataflow perspective, \emph{evaluating} a given
model~$w$ is easily parallelized, since the sum of the losses decompose over
the data points~$(x,y)$.

Example: A regularized linear regression\footnote{a.k.a., linear support vector
regression or ridge regression} is a \emph{linear model}, hence
$f_w(x)=\inner{w}{x}$ is the inner product between the data point~$x$ and a
weight vector~$w$.  Choosing the quadratic distance~$l(f(x),y) =
\frac{1}{2}(f(x) -y)^2$ as the loss function leads to linear regression.
Finally, we select the squared norm of $w$ as the regularizer~$\Omega(w) =
\frac{1}{2}|w|_2^2$:

\begin{equation}
  \hat{w} = \argmin{w} \left(\frac{\lambda}{2} |w|_2^2 + \sum_{(x,y)
      \in D} \frac{1}{2} \left(\inner{w}{x} - y \right)^2\right)
\end{equation}

In most instances, the loss $l(f_w(x),y)$ is \emph{convex in $w$},
which guarantees the existence of a minimizer $\hat{w}$ and
\emph{differentiable}.  This facilitates efficient search strategies
that use the gradient of the cost function with respect to $w$.
Different choices for the optimization algorithm are possible.  Here,
we restrict ourselves to the iterative procedure \emph{Batch Gradient
  Descent} (BGD), as it embodies the core dataflow of a wide variety
of optimization algorithms.  Until convergence, batch gradient descent
performs the following step:

\begin{equation}\label{eq:bgd-step}
  w_{t+1} = w_t -  \left( \lambda\partial_w \Omega(w) + \sum_{(x,y) \in D}   \partial_w l\left(f_w(x),y\right)\right)
\end{equation}

Here, $\partial_w$ denotes the gradient with respect to $w$.  Just as
in the case of evaluating a model~$w$ above, the sum decomposes per
data point~$(x,y)$, which facilitates efficient parallelization and
distribution of the computation of each gradient descent step.

The beauty of this approach lies in its generality: Different choices
for the loss~$l$, the prediction function~$f_w$ and the
regularizer~$\Omega$ yield a wide variety of machine learning models:
Support Vector Machines, LASSO Regression, Ridge Regression and
Support Vector novelty detection to name a few.  All of which can be
efficiently learned through BGD or similar algorithms.

BGD can be captured in Iterative Map-Reduce-Update quite easily.  In fact, the
sum in \eqref{eq:bgd-step} can be efficiently captured by a single
MapReduce step where each \ol{map} task computes gradients for its
local data points while the \ol{combine} sums them up and \ol{reduce}
applies them and the gradient of the regularizer~$\Omega$ to the
current model $w$.  The user needs to supply the UDFs mentioned 
in section~\ref{sec:itermr}:

\begin{description}
\item[\ol{map}] computes a gradient for the current data point, using
  the current model $w_t$
\item[\ol{reduce}] aggregates a set of gradients into one.
\item[\ol{update}] accepts a current model~$w_t$ and the aggregated
  gradients and produces a new predictor $w_{t+1}$ after applying the
  regularizer~$\Omega$.
\end{description}

\section{Model(ing) Semantics}\label{sec:semantics}

Datalog least-fixedpoint semantics tells us that a program without aggregation
and negation has a unique minimal model.  In other words, the result we get
from evaluating the rules to fixpoint is always the same and consistent with
the logic program.  A Datalog program that includes aggregates and negated
subgoals---like those in Section~\ref{sec:declarative}---may have several
minimal models.  There are more general classes of Datalog semantics that can
decide which one minimal model is consistent with the intent of the programmer.
In Section~\ref{sec:stratification}, we show that the programs in
Section~\ref{sec:declarative} are in the class of locally stratified Datalog
programs.  In Section~\ref{sec:termination}, we argue that our runtime
selects the one minimal model that is consistent with locally stratified
Datalog semantics and our conditions for program termination.

\subsection{Program Stratification} \label{sec:stratification}

Stratified Datalog semantics extend least-fixedpoint semantics with a method
for organizing predicates into a hierarchy of strata; using a process called
stratification.  If some predicate $A$ depends on an aggregated or negated
result of another predicate $B$ then $A$ is placed in a higher stratum than
$B$.  A runtime that supports Stratified Datalog evaluates rules in lower
strata first.  Intuitively, this forces the complete evaluation of predicate
$B$ before predicate $A$ is allowed to view the result.  Stratification fails
when there are cycles through negation or aggregation in the (rule/goal)
dependency graph.  Intuitively, if $A$ and $B$ depend on each other, perhaps
even indirectly, then we can not evaluate one to completion while isolating the
other.

Program stratification fails in Listings~\ref{lst:datalog-local} and
\ref{lst:datalog-global} (Section~\ref{sec:declarative}) since they both
contain cycles through a stratum boundary (i.e., aggregation or negation).
Therefore, we look to another class of Datalog semantics called locally
stratified programs, which is defined in terms of a data dependent property.
Intuitively, these programs are not necessarily stratified according to the
syntax of the rules, but rather according to the application of those rules on
a specific data collection.  The following definition follows from Zaniolo et
al.,~\cite{Zaniolo93}.

\begin{dfn}
  A program is locally stratifiable iff the Herbrand base can be partitioned into a
  (possibly infinite) set of strata $S_0$, $S_1$, $\ldots$, such that for each
  rule~$r$ with head~$h$ and each atom $g$ in the body of $r$, if $h$ and $g$ are,
  respectively, in strata $S_i$ and $S_j$, then
  \begin{enumerate}
    \item $i \ge j$ if $g$ is a positive goal, and
    \item $i > j$ if $g$ is a negative goal.
  \end{enumerate}
\end{dfn}

Intuitively, a program is locally stratifiable if the model data---formed from
the initial facts and rule derivations---is stratifiable.  The key to proving
that the programs in Listings~\ref{lst:datalog-local}
and~\ref{lst:datalog-global} are locally stratified lies in the temporal
argument of our recursive predicates.  The values of the temporal argument are
taken from a discrete temporal domain that is monotonic. This allows us to use
another program stratification technique called
XY-Stratification~\cite{Zaniolo93}.

\begin{dfn}
  Let $P$ be a program with a set rules defining mutually recursive predicates.
  $P$ is an XY-Stratified program if it satisfies the following conditions:
  \begin{enumerate}
    \item Every recursive predicate has a distinguished temporal argument.
    \item Every recursive rule is either an X-rule or a Y-rule.
  \end{enumerate}
\end{dfn}

In an X-rule, the temporal arguments of every recursive predicate must refer to
the current temporal state (e.g., $J$).  A Y-rule has the following constraints.
\begin{enumerate}
  \item The head predicate temporal argument value contains a successor state (e.g., $J+1$).
  \item Some positive goal in the body has a temporal argument of the current state (e.g., $J$).
  \item The remaining recursive goals have a temporal argument that contains either the
    current state (e.g., $J$) or the successor state (e.g., $J+1$).
\end{enumerate}
Intuitively, an X-rule reasons within the current state and a Y-rule reasons
from the current state to the next.

It is known that if a program is XY-stratified then it is locally
stratified~\cite{Zaniolo93}.  We now show that the programs in
Section~\ref{sec:declarative} are XY-stratified programs using the following
construction applied to each recursive rule~$r$.
\begin{enumerate}
  \item Rename all recursive predicates that have the same
    temporal argument as the head with a prefix {\bf new\_}.
  \item Rename all other occurrences of recursive predicates with
    the prefix {\bf old\_}.
  \item Drop the temporal arguments from all recursive predicates.
\end{enumerate}
If the resulting program following this construction can be stratified then the
original program is locally stratified~\cite{Zaniolo93}.

\begin{thm}
  Listing~\ref{lst:datalog-global} is in the class of XY-stratified programs.
\end{thm}

\begin{figure}
\centering
\begin{lstlisting}[language=Prolog,
                   caption={Listing~\ref{lst:datalog-global} after XY-Stratification.},
                   label={lst:xydatalog-global}, frame=single]
% Initialize the global model
G1: new_model(M) :- init_model(M).

% Compute and aggregate all outbound messages
G2: new_collect(reduce<S>) :- new_model(M),
  training_data(Id, R), map(R, M, S).

% Compute the new model
G3: new_model(NewM) :-
  old_collect(AggrS), old_model(M),
  old_update(M, AggrS, NewM), M != NewM.
\end{lstlisting}
\end{figure}

\begin{proof}
  The program in Listing~\ref{lst:xydatalog-global} follows from applying
  XY-Stratification to the program in Listing~\ref{lst:datalog-global}.
  Listing~\ref{lst:xydatalog-global} is trivially stratified by placing
  \ol{new\_collect} in the highest stratum.  Therefore, evaluating the rules in
  Listing~\ref{lst:xydatalog-global} produces a locally stratified model that
  is consistent with the programmer's intent in Listing~\ref{lst:datalog-global}.
\end{proof}

\begin{thm}
  Listing~\ref{lst:datalog-local} is in the class of XY-stratified programs.
\end{thm}

\begin{figure}
\begin{center}
\includegraphics[scale=0.43]{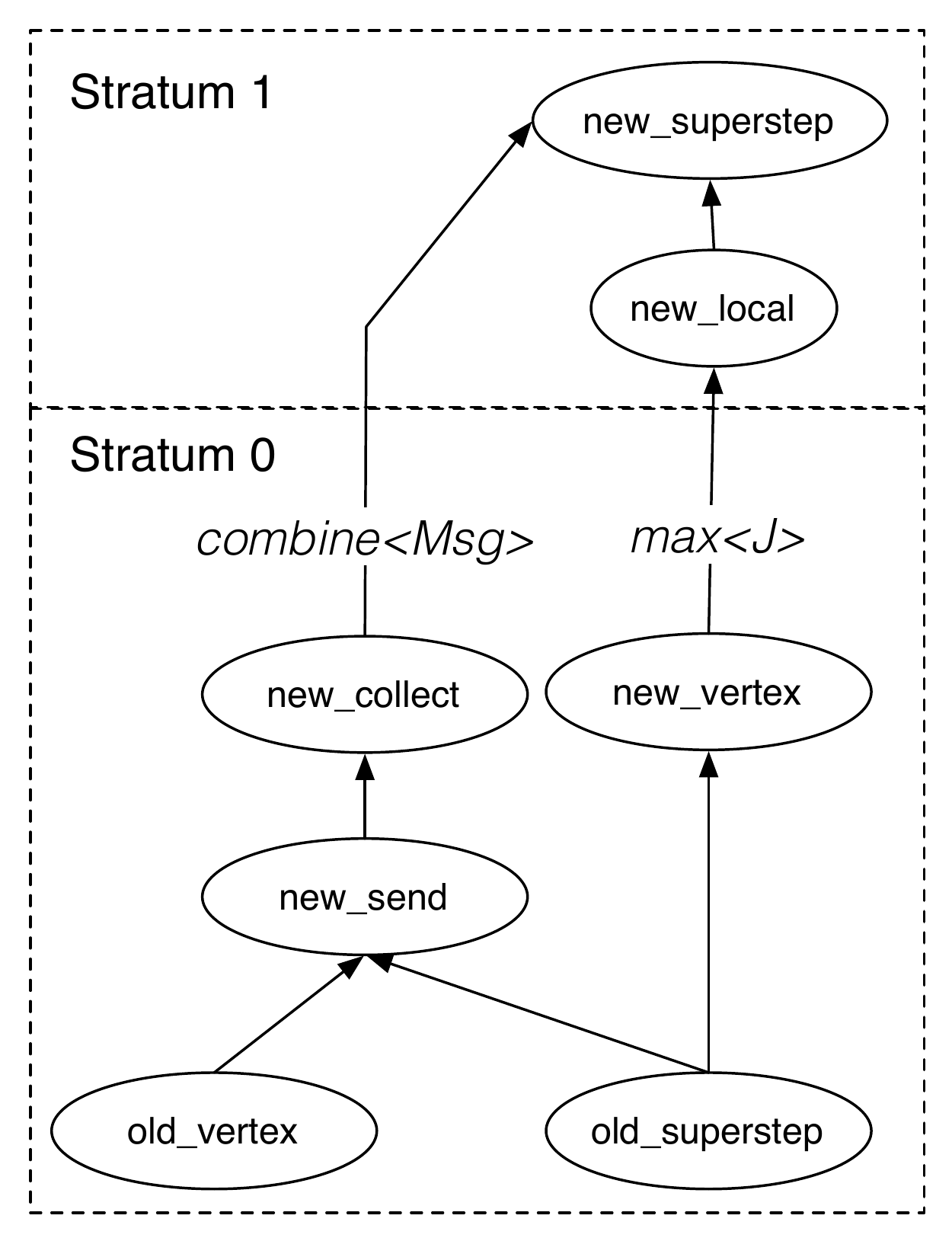}
\caption{\label{fig:dependency-graph}Dependency graph for XY-stratified Listing~\ref{lst:datalog-local}.}
\end{center}
\end{figure}

\begin{proof}
  Figure~\ref{fig:dependency-graph} contains the dependency graph for the predicates
  appearing in Listing~\ref{lst:datalog-local} after the XY-stratified transformation.
  The graph shows that the program is stratified into two strata. We further note that
  naming \ol{new\_local} 
  comes from using \ol{max} aggregation applied
  to the temporal argument of base predicates \ol{new\_vertex},
  and \ol{new\_superstep} comes from using \ol{new\_local} and
 \ol{combine} UDF applied to
 \ol{new\_collect}.
\end{proof}

\subsection{Stratified Evaluation and Termination}
\label{sec:termination}

So far, we have applied XY-Stratification to our programs and to produce new
programs that are stratifiable.  The data in the $i^{th}$ time-step treated
data from previous time-steps $j < i$ as the extensional database (EDB).  This
allowed us to break dependency cycles at Y-rules, which, by definition, derive
data for the subsequent time-step.  These XY-Stratified programs formed the
basis of the template physical plans described in Section~\ref{sec:physical}.

We now conclude with a discussion of termination of our Datalog programs.  The
runtime terminates when the Datalog program reaches a fixpoint.  We have
already shown that the result of a fixpoint is a locally stratified model.
However, this model could be infinite, in which case it would never terminate.
Therefore, termination depends solely on a finite fixpoint solution.  Under
Datalog semantics this occurs when derivations range over a finite domain.
Intuitively, if the range is finite then we will eventually derive all possible
values since Datalog is monotonic and set-oriented.

For the programs listed in Section~\ref{sec:declarative}, this can occur in two
possible ways.  First, when the temporal argument ranges over a finite time
domain.  Since this argument is monotonic and finite, we are guaranteed to
reach an upper bound, and hence terminate.  A second possible termination
condition comes from the range of state values given by the \ol{update}
function.  Recall that this function produces new state objects
when given the (current) state object and list of messages.  The runtime will
consider the \ol{update} function predicate to be false if the new state object does not
differ from the previous.  Therefore, if there are a finite number of possible
state objects, and each state object is produced exactly once, then we are also
guaranteed to terminate.  In other words, there are a finite number of state
objects and the \ol{update} UDF enumerates them in a monotonic
fashion.

\end{appendix}


\end{document}